# A survey on popularity bias in recommender systems

**Anastasiia Klimashevskaia[1] · Dietmar Jannach[1,2] · Mehdi Elahi[1] · Christoph Trattner[1]**

## Abstract

Recommender systems help people find relevant content in a personalized way. One main promise of such systems is that they are able to increase the visibility of items in the *long tail*, i.e., the lesser-known items in a catalogue. Existing research, however, suggests that in many situations today's recommendation algorithms instead exhibit a *popularity bias*, meaning that they often focus on rather popular items in their recommendations. Such a bias may not only lead to the limited value of the recommendations for consumers and providers in the short run, but it may also cause undesired reinforcement effects over time. In this paper, we discuss the potential reasons for popularity bias and review existing approaches to detect, quantify and mitigate popularity bias in recommender systems. Our survey, therefore, includes both an overview of the computational metrics used in the literature as well as a review of the main technical approaches to reduce the bias. Furthermore, we critically discuss today's literature, where we observe that the research is almost entirely based on computational experiments and on certain assumptions regarding the practical effects of including long-tail items in the recommendations.

**Keywords** Recommender systems · Popularity bias · Long tail · Fairness · Diversity

✉ Anastasiia Klimashevskaia
  Anastasiia.Klimashevskaia@uib.no

  Dietmar Jannach
  Dietmar.Jannach@aau.at

  Mehdi Elahi
  Mehdi.Elahi@uib.no

  Christoph Trattner
  Christoph.Trattner@uib.no

[1] MediaFutures: Research Centre for Responsible Media Technology & Innovation, University of Bergen, Bergen, Norway

[2] AAU Klagenfurt, Klagenfurt am Wörthersee, Austria

Springer

# 1 Introduction

Recommender systems are nowadays used by many online platforms—including most major e-commerce and media streaming sites—where they can create substantial value for both consumers and providers (Jannach and Zanker 2021). From the consumers' side, these systems, for example, may support them in finding relevant content in situations of information overload or help them *discover* the content that was previously unknown to them. On the provider's side, on the other hand, recommendations can effectively improve engagement, stimulate cross-sales or help promoting items from the *long tail* (Anderson 2006) of less popular and probably hard-to-find items. Among the various possible benefits of recommender systems, they seem to be particularly suited to support a long tail business strategy. By surfacing more of the long tail items in personalized way, they support both the goals of improved *discovery* of new content for consumers as well as increased benefit for the provider, e.g., in terms of increased user engagement, customer retention, additional sales or changed demand curves, see Celma (2010), Gomez-Uribe and Hunt (2015), Oestreicher-Singer and Sundararajan (2012).

While there is no doubt that recommender systems can effectively impact consumer behavior and shift sales distributions (Lawrence et al. 2001; Zanker et al. 2006), it turns out that in practical settings such systems can have unexpected effects. For instance, the results of a large-scale field test on a North-American retailer site revealed that a recommender system indeed has had a positive effect on the sales of niche items. However, the increase that was observed for the popular items was even more pronounced. Moreover, aggregate sales diversity actually *decreased* in the presence of the recommender (Lee and Hosanagar 2014, 2019). Such observations can be attributed to a certain *popularity bias* in the underlying algorithms, which means that the algorithms may have a tendency to focus on already popular items in their recommendations. As a result, the already popular ("Blockbuster") items (Fleder and Hosanagar 2009) receive even more exposure through the recommendations, which can ultimately lead to a feedback loop where the "rich get richer".

Overall, a too strong focus on popular items can be disadvantageous both for consumers and providers in all sorts of application domains of recommender systems. Consumers might find the recommendations obvious, not novel enough, and thereby not supporting the need for discovery. Providers, on the other hand, not only fail to supply adequate discovery support, but also miss the opportunity to sell from the long tail by mainly promoting items which customers might have bought or consumed anyway (Bodapati 2008). Given the high practical importance of the problem, an increasing number of research works have addressed the problem of popularity bias in recommender systems over the last decade. In particular, in most recent years the topic has become prevalent in the light of fairness and biases in recommender systems (Chen et al. 2020; Ekstrand et al. 2022), as well as in the context of potential harmful effects of recommendations such as filter bubbles, echo chambers, persuasion and manipulation (Aridor et al. 2020; Elahi et al. 2021a).

With this paper, our goal is to provide a multi-faceted overview on the current literature on popularity bias in recommender systems, a topic that has drawn considerable attention in recent years. To that purpose, we systematically reviewed and categorized 123 papers along various dimensions, e.g., according to the underlying research motivations, the technical approaches to deal with popularity bias, and evaluation methodologies. Among other aspects, the following key insights emerged from our analyses.

– In terms of underlying research motivations, most examined works are based on the application-*independent* assumption that focusing on popular items is problematic *per se* and causes potential consequences such as limited exposure of certain items, the reinforcement of biases, and limited recommendation quality for users by default. Much of the current literature also seems to be fueled by the growing interest in fairness in recommender systems. Application-*specific* considerations, e.g., at which point popularity bias may actually be harmful or lead to unfairness in a given context of use are mostly missing. With our work, we aim to provide a more nuanced discussion of the consequences of popularity bias, and offer a novel, impact-oriented definition of popularity bias.

– A rich variety of computational methods are proposed in the literature, mostly to quantify the existing bias or to mitigate it. The mitigation approaches themselves can be categorized as pre-processing, in-process (modeling), and post-processing approaches. We find that in-process techniques, which support the joint consideration of competing objectives, are the most common form of mitigating bias in the literature.

– From an evaluation perspective, we observe that the literature is heavily relying on offline experiments and a rich variety of abstract and application-independent computational metrics. Studies with users or field tests are very rare. This phenomenon, like in the area of research in fairness in recommender systems and fairness in AI in general, may lead to a certain 'abstraction trap' (Selbst et al. 2019), where the operationalization of the research problem abstracts too much of the idiosyncrasies of specific application use cases. This, as a result, may lead to a certain gap between academic research and real-world problem settings.

The rest of the paper is organized as follows. We first elaborate on existing definitions of the concept and possible sources of popularity bias in Sect. 2. After describing our research methodology to identify relevant papers in Sect. 3, we provide statistics regarding the different types of contributions we observe in the literature in Sect. 4. We discuss technical proposals to deal with popularity bias in Sect. 5 and we review evaluation approaches in Sect. 6. The paper ends with a discussion of our insights and an outlook on research gaps and possible future directions in Sect. 7.

## 2 Background

In this section, we define the term popularity bias, discuss the possible sources of bias in more depth, and outline practical negative effects resulting from popularity bias.

## 2.1 Popularity bias as an exposure-related phenomenon

While we observe a largely shared understanding in the research community regarding the potential harms of popularity bias in recommender systems, no unique definition seems to exist so far. Most commonly, popularity bias is considered a *characteristic of the recommendations* that are shown (exposed) to users.

In Abdollahpouri and Mansoury ([2020](#)), for example, popularity bias is described as a phenomenon where "*popular items are recommended even more frequently than their popularity would warrant.*" In such an interpretation, the bias exists when the system recommends popular items to an exaggerated extent. Similar considerations regarding disparities in the recommendations were discussed in other works as well, e.g., in Lesota et al. ([2021](#)). In other definitions, however, such proportions are not in the focus, and an emphasis on popular items per se is considered a bias. According to Abdollahpouri et al. ([2017a](#)), "*collaborative filtering recommenders typically empha-size popular items (those with more ratings) much more than other 'long-tail' items.*" Similarly, Boratto et al. ([2021](#)) state that popularity bias can be described as the effect that recommender systems may "*tend to suggest popular items more than niche items, even when the latter would be of interest.*" Such a concept is also adopted in Zhu et al. ([2021a](#)) and other works.

We note that Boratto et al. in their discussion connect the bias that is observed in the recommendations with an underlying reason, i.e., the bias occurs when algorithms are trained on datasets where the observed interactions are not uniformly distributed across items. In some works, such skewed distributions themselves are referred to as popularity bias, thus framing popularity bias as a *characteristic of the training data* that a recommender system picks up on. Zhao et al. ([2022](#)), for example, found that "*the observation data usually exhibits severe popularity bias, i.e., the distribution over items is quite imbalanced and even long-tailed.*"

Finally, some works discuss popularity bias in recommender systems in the context of offline evaluation metrics. A particular challenge in this context can be that certain metrics, and in particular *precision*, can favor algorithms that have a tendency to recommend popular items. By averaging across users, optimizing for high precision means to try to satisfy the majority of the (popularity-oriented) users, "*regardless of the satisfaction of minorities*" (Bellogín et al. [2017](#)). This may then lead to a competitive performance of non-personalized and popularity-oriented methods (Cremonesi et al. [2010](#)), and alternative evaluation protocols are proposed to deal with such problems, see also Bellogin et al. ([2011](#)), Bellogín et al. ([2017](#)), Ekstrand et al. ([2018](#)), Mena-Maldonado et al. ([2020](#)), Yang et al. ([2018b](#)).

In this work, we adopt the previously discussed viewpoint and terminology where popularity bias is a phenomenon that is related to the popularity of the items that are recommended to users. Thus, we separate the observed phenomenon from the potential underlying sources of popularity bias.

## 2.2 Sources of biases and bias amplification

In most research works on recommender systems, the popularity of an item is assessed by the number of user interactions (e.g., ratings, clicks, purchases) that are observed in a dataset. We note that in most applications of recommender systems we actually would not expect a balanced distribution. In many domains, there may be items that are more popular than others. Some products in an e-commerce store might, for example, be of better quality or cheaper in price than others or strongly promoted through advertisements, leading to more observed purchases. In the entertainment domain, on the other hand, some movies or musical tracks may just appeal to a broader audience and we may therefore record more streaming events. We refer to such pre-existing, commonly skewed distributions regarding the popularity of items as the *natural bias* in the data.

Besides this potentially pre-existing bias in the data, we note that additional bias can be introduced when selecting the data to be used for training a recommender system (*data collection or representation bias*). For example, only a certain subset of the user base might be considered when creating a training dataset, but this subset may not be fully representative of the entire population.

In any case, while at least parts of a given imbalance in a collected dataset may appear natural, a serious problem of recommender systems is that they might *reinforce* these pre-existing distributions. Ultimately, this reinforcement may lead to detrimental effects in the long run, where the system increasingly puts more emphasis on already popular items, thereby reducing the chances of lesser known items to be exposed to users. Chen et al. (2020) identify various factors that may ultimately lead to a feedback loop in recommender systems, as shown in Fig. 1.

Internally, many recommender systems these days are based on some type of machine learning model. A central ability of any machine learning algorithm is to generalize from past experience (training instances) to deal with new situations (unseen instances) (Mitchell 1990). Therefore, what an algorithm learns always reflects to a certain extent what is observed in the training data, including in particular any (pre-existing) bias in the data. We note here that each algorithm may have its own inductive biases, i.e., a set of assumptions when performing the inductive leap from the training data to the general model (Hüllermeier et al. 2013).

Let us consider the very basic scenario of recommending shopping items that are *frequently bought together*, as implemented in today's major e-commerce platforms. Technically, recommendations of this type can be seen as a basic form of *association rules* (Agrawal et al. 1993; Ludewig et al. 2021). A common challenge in the rule mining task is that the rules with the highest support commonly involve very popular items, and that it is challenging to determine rules that involve niche items (Uday Kiran and Krishna Re 2009). Thus, it is intuitive to assume that item suggestions that are based on a "frequently-bought- together" statistic have a tendency to further reinforce the promotion of already popular items.

More generally, the suggestions that are subsequently made to users based on a machine learning model reflect to a certain extent what the recommender system has learned from the data and how it was optimized. In particular, depending on the optimization metric during training, the algorithm may have learned—although not necessarily explicitly—that recommending popular items will give high "reward" in terms of the metric.

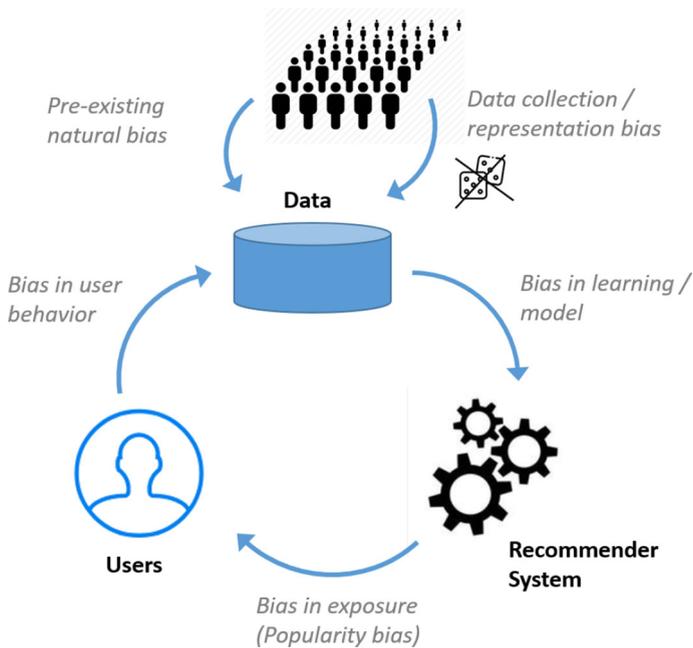

**Fig. 1** Biases and the feedback loop of recommendation, inspired by Chen et al. (2020)

Ultimately, the recommendations presented to users are generally assumed to be able to influence their choices to a certain extent. Higher-ranked items in recommendation lists commonly receive more exposure and user attention and, consequently, are more likely to be consumed (Joachims et al. 2007), e.g., due to position bias. As a result, they may be consumed or purchased more often than other options. Thus, in case where the recommendations are influenced by popularity bias, it finally means that the already popular items profit more from this increased exposure than some lesser known ones. Importantly, when users adopt (i.e., consume or purchase) a recommended popular item, this fact will commonly be reflected in some ways in the data that is used to retrain the underlying model in a subsequent step. A successful recommendation of a popular item will, for example, further increase an item's purchase statistic. Moreover, as popular items are often good recommendations in terms of their general quality and appeal, the chances that they receive positive feedback, e.g., in the form of a rating, may also be high if we assume that people tend to provide feedback on things that they like. This corresponds to the known problem that certain data points are "missing-not-at-random", see Marlin et al. (2007) for an early study on the topic.

Overall, we observe that there are various stages where popularity bias can enter or be reinforced in a recommender system. Correspondingly, different approaches and starting points exist when the goal is to mitigate the potentially undesired effects of popularity bias in the recommendations.

## 2.3 Potential negative effects of popularity bias

Research on popularity bias is commonly motivated with examples of possible negative effects when an algorithm focuses (too much) on already popular items. Sometimes, recommending popular items is considered problematic, as this may *unfairly* reduce and prevent the exposure of other items. In other cases, reference is made to potential reinforcement effects over time, often circumscribed as a situation where the "rich get richer", a phenomenon which is sometimes referred to as "Matthew Effect" (Wang et al. 2018) or "Prefix Bias" (Rashid et al. 2002).

At first sight, one may argue that there is nothing wrong with recommending popular items. In fact, recommending top selling items is quite common also in the offline world, e.g., in the form the *New York Times Best Seller* book recommendations. Moreover, in a meritocratic society, it may not be considered problematic or unfair if these best sellers receive even more attention through recommendations, assuming that they are of higher quality than others or generally appealing to more people. As such, the above mentioned claims about potential harms of popularity bias sometimes seem too general.

However, when looking closer at the problem and the intended purpose and value of a recommender system (Jannach and Zanker 2021), one can easily derive a number of ways in which popularity bias *(a)* either *limits the potential value* of the recommendations for individual stakeholders or *(b)* where the *bias may actually be harmful*. In terms of limited value, consumers may find that popularity-biased recommendations do not help them to *discover* new content (because of limited novelty) or content that matches their personal preferences (because of a limited level of personalization). Both aspects may in turn limit the engagement of consumers with the service that provides the recommendations or turn them away completely by losing their trust in the system. On the provider's side, recommending mostly popular items may furthermore lead to *missed sales opportunities* (because the popular items would have been purchased anyway). Moreover, it may lead to decreased sales diversity over time (because a small set of popular items receives all the exposure). Corresponding reports from field and simulation studies can be found in Fleder and Hosanagar (2009), Jannach et al. (2015), Ferraro et al. (2020).

Situations where a popularity-biased system may actually create harm (and not only provide limited value) can also arise in certain application domains. In recent years, various research works on fairness in recommender systems—see Ekstrand et al. (2022), Wang et al. (2022b), Deldjoo et al. (2023) for recent surveys—argued that popularity bias can lead to unfairness. For example, certain jobs may be mainly recommended to particular ethnic groups when the recommender systems perpetuates historical discrimination. Alternatively, a music recommender system may unfairly mostly promote music from certain groups of already popular artists, limiting the chances of exposure for artists which, e.g., may belong to an underrepresented gender or genre groups.

Another, yet quite different, harmful case may occur when a popularity-biased system promotes content that is harmful. We recall that in many applications popularity is measured in terms of the observed interactions with an item. In particular, in social media it is not uncommon that controversial content (including fake news, misinformation and disinformation) receives a lot of attention as users are highly engaged with such content. A social media recommender system that optimizes for user engagement may therefore further promote such questionable content by suggesting it to an increasingly larger audience. Furthermore, such a popularity-biased system may also be vulnerable to recommend content which received many interactions through fake users, false reviews/ratings and automated bots, see, e.g., Lam and Riedl (2004), which may ultimately lead to a loss of trust in the system.

Overall, regardless of whether the utility is reduced or actual harm is caused, it is important to consider the specifics and idiosyncrasies of a particular application use case when investigating questions of popularity bias. On the one hand, recommending popular items can in fact be the most beneficial option for a provider, e.g., when the top-selling items are also the ones that lead to the highest revenue, profit margin or other business Key Performance Indicator (KPI). On the other hand, recommending already popular items should not be considered unfair per se, but one has to scrutinize which underlying normative claims regarding fairness are affected by popularity-biased recommendations. Furthermore, we have to keep in mind that certain effects may only become visible in the long term. Promoting the most popular and recent celebrity gossip on a news website might lead to positive effects in the short run in terms of the click-through rates (CTR); it may however lead to limited engagement with the service in a longitudinal perspective.

Finally, we note that focusing on popular items can be a beneficial and helpful approach as well in certain situations. Recommending popular items is a very common strategy in *cold-start* situations where little is known about the preferences of the user. For example, when a new user registers to a recommender system, the system has no or limited knowledge about the user's preferences and hence may fail to generate relevant recommendations for her. In such a case, a popularity-based *active learning* strategy can be employed to select the top popular items to be proposed to the new user and acquire explicit ratings for them Rashid et al. (2002). The advantage is that the user is very likely to be familiar with the popular items and hence can actually be able to rate these items. Despite the positive side, popular items are typically liked by the users and hence, their ratings often bring little information to the system (Elahi et al. 2016).

Furthermore, there can be situations where a specific algorithm focuses too much on niche content. In such cases, the recommendations might appear too obscure for users, not raise their interest, and limit their satisfaction with the service (Ekstrand et al. 2014). Including a number of popular recommendations may help establish a certain level of familiarity with the recommendations at the user's side, and their trust that some recommendations are suitable for them. Adding a "*healthy dose of (unpersonalized) popularity*" is also not uncommon in industrial settings, e.g., for the personalized video ranking system at Netflix (Gomez-Uribe and Hunt 2015).

### 2.4 An impact-oriented definition of popularity bias and its relationship to novelty, diversity, and fairness

As discussed in the beginning of this section, there is no unique definition of the term popularity bias in the literature. Some definitions may also be not easy to interpret or apply. If we, for example, develop a recommender system that simply recommends the most popular items to everyone, it may be difficult to tell if this would represent a case where items are recommended "*more frequently than their popularity would warrant*", as described in Abdollahpouri and Mansoury (2020) or Chen et al. (2020). Moreover, our discussions also show that recommending popular items is not necessarily harmful per se, and that it instead may depend on the particularities of a given use case.

Following our discussions and under the assumption that the term *bias* generally indicates an undesirable or problematic aspect, we propose to use an *impact-oriented* interpretation of the term in the future. Accordingly, we propose to define popularity bias in recommender systems as follows.

> *A recommender system faces issues of popularity bias when the recommendations provided by the system focus on popular items to the extent that they limit the value of the system or create harm for some of the involved stakeholders.*

We emphasize that our definition is aimed to be generic and encompassing in the sense that it does *(a)* not prescribe a specific way in which popularity is quantified, *(b)* it does not make assumptions about the sources of the bias, and *(c)* it may include both short-term or long-term effects of popularity bias.

The popularity of the recommended items is related with a number "beyond-accuracy" quality aspects of recommender systems, in particular to novelty, diversity, and serendipity (Castells et al. 2021; Kaminskas and Bridge 2016; Ziarani and Ravanmehr 2021).

*Relationship to Novelty.* A recommendation provided to a user is usually considered to be novel if the user has not previously known about it Castells et al. (2021), Kaminskas and Bridge (2016). Novelty is thus a central desirable feature, as novel recommendations per definition help users discover new (and hopefully relevant) things. The perceived novelty of a set of recommendations can be empirically assessed with the help of user studies (Ekstrand et al. 2014; Pu et al. 2011). In offline evaluations, we, in contrast, often cannot know with certainty if a user already knows an item. A common approach in the literature, therefore, is to assume that less popular items, on average, have a higher probability of being novel for the users. Technical realizations of novelty metrics are therefore frequently formulated as being inversely related to popularity metrics (Vargas and Castells 2011). Typically, a common goal in novelty-focused research is to increase the novelty level (or: reduce the popularity level) of the recommendations without sacrificing accuracy. In such settings, novelty-enhancing approaches can also be seen as methods to decrease popularity bias.

Serendipity is another concept that is related to novelty. Often, serendipity is viewed as a combination of unexpectedness and relevance (Ziarani and Ravanmehr 2021), but other notions exist as well in the literature (Ziarani and Ravanmehr 2021). Clearly, a serendipitous item must also be novel. However, an item is often only considered unexpected if it is in some ways different from a user's usual taste profile.

We note here that *item discovery*, as supported through novel or serendipitous item recommendations, is one of the most common purposes of a recommender system. However, also use cases exist, where a recommender system explicitly aims to suggest already known items, e.g., to stimulate repeated purchases in an e-commerce setting or to remind users of previously liked content on a streaming platform (Kapoor et al. 2015; Lerche et al. 2016).

*Relationship to Diversity.* Diversity often refers to the property that the elements of a set of recommendations differ from each other in certain aspects (Ziegler et al. 2005; Kaminskas and Bridge 2016). Depending on the selected criterion and use case, popularity bias *can* be related to diversity. In certain domains, e.g., in movie recommendation, suggesting widely known popular movies will probably result in a set of movies that is not too diverse in terms of the country of the production, the production budget, or the original language. If the popularity level of these recommendations is decreased, we may therefore observe an increase in diversity in these aspects.

Other notions of diversity include *sales diversity* (Fleder and Hosanagar 2009), which measures the concentration of the sales volume on certain items, or *aggregate diversity* (Adomavicius and Kwon 2011), which is a sort of coverage metric that measures the fraction of catalog items that are recommended to users in *top-n* lists. In the case of aggregate diversity, a stronger focus on mostly popular items leads to a lower level of personalization, and, expectedly, to a more limited catalog coverage. A field study on the effects of recommender systems on sales and *sales diversity* (Lee and Hosanagar 2019) however led to partially unexpected results. First, it was observed that implementing a recommender system led to a *decrease* in sales diversity, which in a way confirms the assumption that recommender systems lead to a concentration effect which may reinforce popular items. In terms of absolute sales, the recommender led to an increased sales volume for long-tail items, which is one expected benefit of recommender systems in the fist place. However, an even stronger increase in sales was observed for popular items.

Overall, we conclude that popularity bias may impact diversity. The relationship is, however, not so direct as for the case of novelty, and the observed effects depend on the particular notion of diversity.

*Relationship to Fairness.* Quite a number of recent research works equate the reduction of popularity bias with an increase of algorithm fairness, see Deldjoo et al. (2023). Certainly, there may be use cases where this may be true. For example, there might be a group of artists on an online music platform which for societal or historical reasons do not have the same opportunity to reach a broad audience as others, e.g., because they belong to a generally underrepresented group. A recommender system that gives more exposure to the less popular content by these artists may then be considered to support a normative claim regarding the fairness towards the underrepresented group (Dinnissen and Bauer 2023; Ferraro et al. 2021). This latter aspect of addressing an underlying normative claim is however essential. Simply increasing the exposure of arbitrary artists on a music platform or the exposure of items of certain providers on an e-commerce platform does not necessarily serve a fairness goal. In fact, some items may just not be popular because they are not generally appealing to a broader audience or because they are of limited quality.

Besides leading to unequal exposure, as mentioned above, popularity bias can have other negative fairness-related effects. For example, the existence of popularity bias can cause inconsistency in the performance of recommender systems when serving different groups of users. Such inconsistency may lead to discrepancies in the recommendation quality (Yao and Huang 2017). This can be interpreted as a form of *miscalibration* in the performance of the system and can be observed as evidence for unfairness in the representations of the interests of users in different groups (Abdollahpouri et al. 2020b; Ekstrand et al. 2018).

In sum, we can conclude that popularity bias *can* have negative effects on different notions of fairness. Reducing popularity bias, however, does not necessarily improve fairness in general, as it depends on the particular underlying normative claim that is connected to a particular fairness consideration.

*Discussion.* Overall, we find that popularity bias can impact various aspects of recommendation quality (including accuracy, diversity, novelty or fairness), and it can lead to possibly undesired effects from an organizational perspective, like decreased sales diversity. In this work, we primarily concentrate on existing application-independent technical approaches to quantify and mitigate popularity bias. Therefore, our focus is not so much on the various possible interactions of popularity bias with other quality aspects, in particular as these interactions may strongly depend on application-specific aspects.

## 3 Methodology

*Paper Retrieval Method.* We adopted a semi-systematic approach to identify relevant research works. In our approach, we applied principles of systematic reviews as discussed in Kitchenham (2004), but we also relied on additional means to discover additional papers in this constantly developing area. The overall process is illustrated in Fig. 2.

In the first step, we queried digital libraries to find an initial set of works on recommender systems published between January $1^{st}$ 2000 and January $31^{st}$ 2024 that have the terms "popularity bias" and "recommender / recommendation / recommendations" in the abstract or keywords. Looking these terms up in paper titles has proven to be too narrow of a query, and at the same time searching through the text of the paper itself has returned too many irrelevant works that just barely mention "popularity bias" as possible related topics. Thus, we concluded that searching through abstracts and keywords should be the most precise method. We used the following query term: "*popularity bias*" AND ("*recommender*" OR "*recommendation\**").[1] The search was last executed on January $24^{th}$, 2024, and the processes returned 129 papers.

Next, we applied a snowballing procedure to identify more relevant works by following the references cited in the initial set of works. Furthermore, we used the *Connected Papers* online tool[2] to find additional related works, also using the keyword

---

[1] The specific syntax is different for the used libraries. As digital libraries, we considered the ACM Digital Library, SpringerLink, ScienceDirect and IEEE Xplore.

[2] https://www.connectedpapers.com.

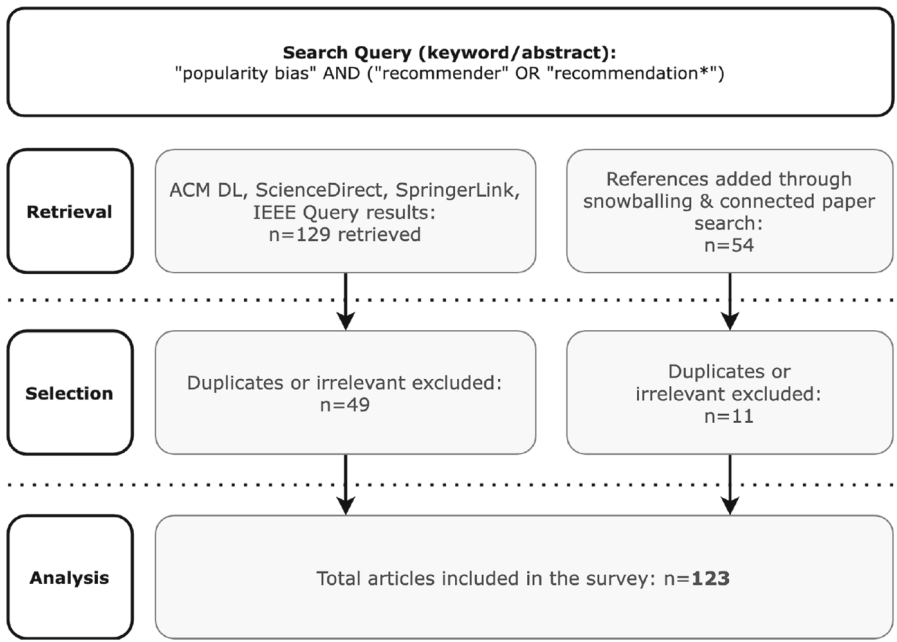

**Fig. 2** Literature collection methodology

"long tail". After removing duplicates and filtering out works which were irrelevant to our survey in a manual process, we ended up with 54 papers, which we considered for the subsequent analyses in our study. We share the detailed list of the considered papers online for reproducibility.[3]

Generally, our search query turned out to be quite precise and the large majority of papers that were retrieved through the query were relevant. There were only a few papers which we considered not relevant. These were papers whose main research contribution was not about popularity bias. For example, some works mentioned the term popularity bias somewhere in the text—which is why they were returned by our search—but then provided a technical contribution that focuses on a different aspect, such as accuracy. Furthermore, we did not consider existing survey works for our analysis.

*Relation To Other Surveys.* The topic of popularity bias has been considered previously in surveys on related topics, such as biases in recommender systems, in general (Chen et al. 2020), undesired effects of recommender systems (Elahi et al. 2021a), or fairness issues in recommender systems (Abdollahpouri et al. 2020a). While our work overlaps with these works to a certain extent, our study is exclusively focused on the problem of popularity bias. Considering the influential survey presented by Chen et al. (2020), for example, we find that this other survey is much broader in scope than ours. They, for example, explicitly include the term 'fairness' in their search query. Moreover, various types of bias in the feedback loop are discussed in Chen et al. (2020), e.g., user conformity bias. Given this breadth of their scope, different technical approaches to specifically deal with popularity bias are not discussed in great depth. Our work, in contrast, aims to provide an in-depth coverage of the topic of popularity bias, with a focus on technical approaches and a survey of common evaluation methodologies.

---

[3] https://docs.google.com/spreadsheets/d/1lvLtrlItfHyxwfc4GzUX-6aVR6rChq3WsbMO9dBVyK4/.

To our knowledge, a recent conference paper (Ahanger et al. 2022) is the only work that exclusively focuses on popularity biases in recommender systems. In their paper, the authors report the technical details of a selected set of recent algorithmic approaches to mitigate popularity biases. While our work is also concerned with technical approaches to bias mitigation, the scope of our present work is broader and we also aim to reflect on the developments in the area. Moreover, differently from this previous survey, our work is based on a larger collection of research works which we retrieved through a structured process as described above.

## 4 Survey results: a landscape of research

In this section, we will first provide more statistics about publication outlets and the interest in the topic over time. Next, we will paint a landscape of existing research in terms of how scholars characterize the problem and what kind of contributions we can find in the literature.

### 4.1 Publication statistics

The earliest paper considered in our study was published in 2008. We note that this paper was not explicitly using the term "popularity bias", but it focused on how to deal with less popular items from the long tail in recommender systems (Park and Tuzhilin 2008). During the next few years, only a few relevant papers were found. Since around 2018, however, we observe a strong increase in the research interest in the topic, in particular also using the term "bias". We may assume that much of the recent research in this area may also be fueled by the growing awareness and interest in the topic of fair recommendations, see Wang et al. (2022b). As a result, a large majority (around 70%) of the considered works were published in the last five years.

Figure 3 shows where the identified research works on popularity bias were published. We collapsed all outlets into the categories "Other" for cases where we found only one single relevant paper for this outlet. There were as many as 48 outlets of that type. These 48 outlets are quite diverse in different dimensions. They include both computer science journals with a rather broad scope as well as rather focused ones, e.g., on machine learning and its applications. Also, the outlets comprise both long-established, prestigious journals as well venues of somewhat lower visibility and reach. Overall, we considered 65 different outlets where papers on popularity bias were published. While over dozen papers were published at ACM RecSys, the most important outlet in this survey, the figure shows that research on the topic is highly scattered. This emphasizes the need for a survey as presented in this paper.

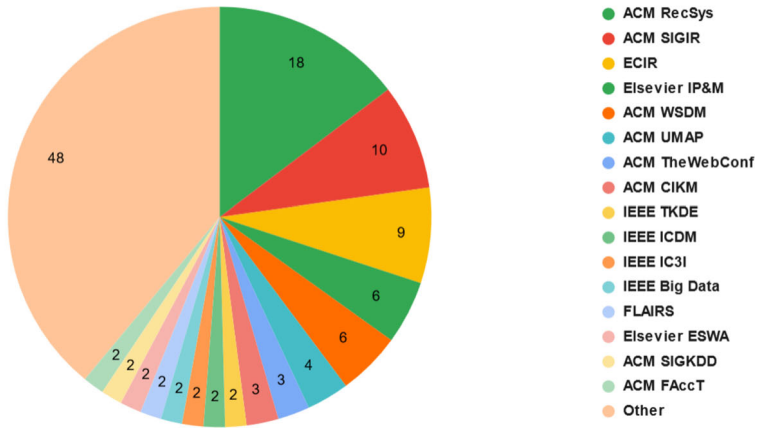

**Papers per outlet**

- ACM RecSys
- ACM SIGIR
- ECIR
- Elsevier IP&M
- ACM WSDM
- ACM UMAP
- ACM TheWebConf
- ACM CIKM
- IEEE TKDE
- IEEE ICDM
- IEEE IC3I
- IEEE Big Data
- FLAIRS
- Elsevier ESWA
- ACM SIGKDD
- ACM FAccT
- Other

**Fig. 3** Number of papers per outlet. Outlets grouped under the label "Other" each have one published work included in this survey

## 4.2 Problem characterizations and research motivations

Following our discussions above, recommending popular items may not be problematic *per se*, and in practice one has to take into account the specifics of the given use case, for example, to determine the extent to which a given bias should be mitigated.

In the first step of our analysis, we investigated how researchers motivate their work. To that purpose, we scanned all papers for statements in the abstract and introduction that characterize the phenomenon of popularity bias as well as the potential harms of recommending popular items. We then applied a coding procedure to identify different categories of such statements. The coding was done by two researchers.

Figure 4 shows along which themes researchers *characterize* the phenomenon of popularity bias. We note that individual papers can fall into more than one category. In the majority of cases, researchers mainly state in some form that popularity bias mainly or too strongly focus on popular items in the recommendations. This generally matches a central part of our definition from the previous section, i.e., that popularity bias is a phenomenon related to the recommendations that are presented to users. Only a comparably small number of papers characterize popularity bias as a phenomenon of the underlying data. However, many papers which rely on such a characterization implicitly assume that focusing on popular items is considered problematic in itself, which may represent an oversimplification of the problem.

The second most frequent characterization is that in the presence of popularity bias, long-tail items receive too limited exposure. While in some sense this might be seen as a direct consequence of the previous aspect, i.e., that a system may focus too much on popular items, this characterization also points to a potential harm, which is a crucial aspect according to our definition. However, only a few works mention that popularity bias may hinder the recommendation of *relevant* long-tail items, which in reality is a highly crucial aspect. A few other works consider questions of recommendation quality in their characterization. In a few cases, popularity bias is assumed to lead to better predictions for popular items. Other works in some ways fear quite the opposite, i.e., that the bias leads to the recommendation of irrelevant popular items.

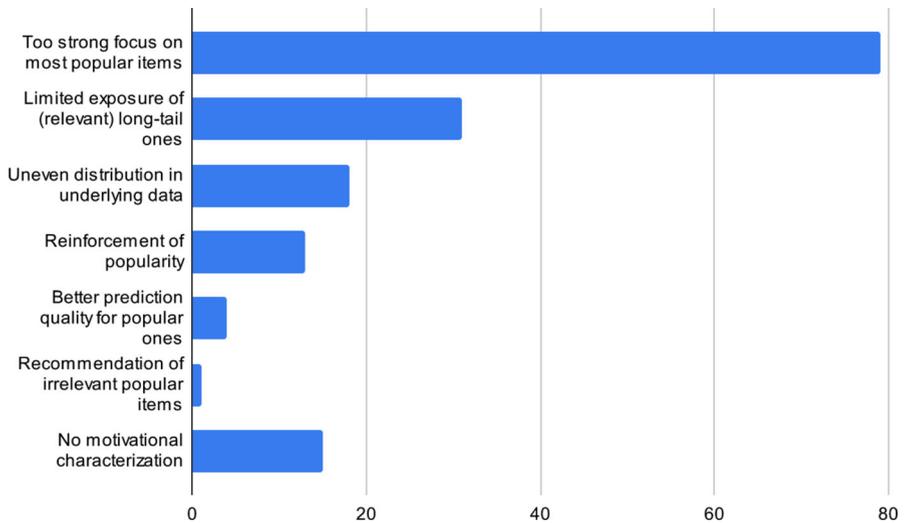

**Fig. 4** Problem characterizations in the literature

Potential reinforcement effects are mentioned a number of times as a main aspect of popularity bias. However, when considering the technical contributions and experimental evaluations provided in many of these papers, the reinforcement effect is not actually investigated, e.g., by assessing the effect from a longitudinal perspective.

Finally, in a certain fraction of papers, we could not identify a clear motivational characterization of the investigated problem of popularity bias. Such papers for example analyze relationships between different quality metrics for recommender systems (including the popularity of the recommendations), without elaborating in depth about the underlying concept, e.g., Channamsetty and Ekstrand (2017). Others like Wu et al. (2019) consider skewed data distributions in their algorithmic design as one of several aspects. Finally, some works like Deldjoo et al. (2021) provide a formal definition for a particular notion of popularity bias, but consider popularity bias as one of several variables in a quantitative analysis of recommendation performance.

Next, we scanned the abstract and introductions for statements that describe the potential *negative effects* of the bias. Such a description of the negative effects should generally guide the research presented in the paper, e.g., in terms of the evaluation metrics. The results of the coding process are shown in Fig. 5.

Some of the most frequently mentioned harms refer to the *recommendation quality* as experienced by the users. Popularity bias may manifest itself in limited personalization quality, limited diversity or novelty, or in terms of limited opportunities for discovery. However, there is also a significant number of works which mention potential harms for the recommendation platform or the item providers, including limited exposure of certain items, missed business opportunities, or reduced consumer trust over time. Some works also raise the issue of potential vulnerabilities in terms of attacks on recommender systems which consider item popularity as a main factor to rank items highly. These observations clearly indicate that there is awareness in the community that popularity bias is a problem that may affect multiple stakeholders. We will discuss later in Sect. 10 how researchers quantify to what extent algorithmic approaches may help to reduce or prevent potential harms of popularity bias.

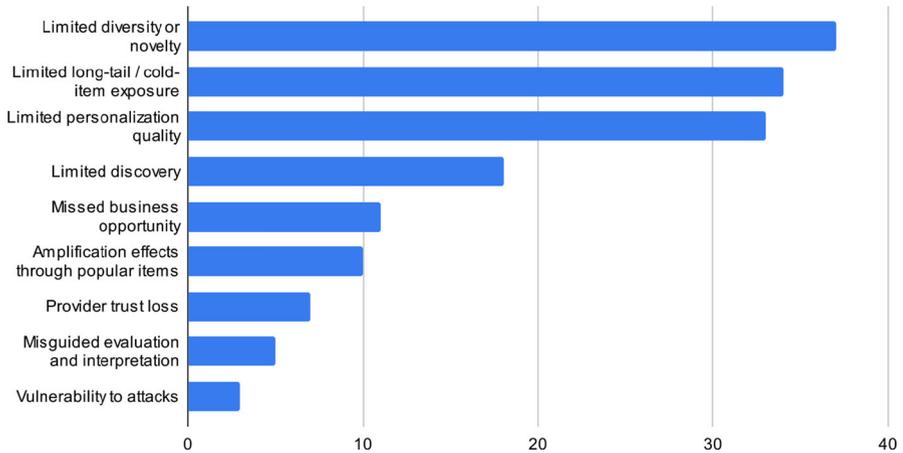

**Fig. 5** Researcher motivation: potential negative effects

Given the recent interest in the community on questions of fairness of recommender systems, we finally scanned the descriptions of potential harms that we found in the paper for the term 'fair'. Only a few works explicitly mention fairness or unfairness in this context, specifying what their definition of fairness is and who they are targeting as a stakeholder. However, considering the broader research setting addressed in the papers, we found that 56 of the 123 papers (about two thirds) do address questions of fairness in recommender systems. This confirms our intuition mentioned above that research on popularity bias in recommender systems is largely fueled by recent fairness research. Again, given that recommendation is a multistakeholder problem (Abdollahpouri et al. 2020a), different forms of fairness are considered in the examined works, including user fairness, item fairness, and provider fairness, see Burke (2017). A slightly larger fraction (60%) of these works focus on user fairness, while the remaining works consider the perspective of items and their providers.

### 4.3 Application domains

Figure 6 provides an overview on the application domains that are considered in the examined works. The application domains were mainly identified based on the datasets that are used in the offline experiments. Similar to other survey works, e.g., Quadrana et al. (2018), we grouped datasets into higher-level categories as shown in Fig. 6.

We can observe that the large majority of works focus on the *media* domain, including movies, music, books, and news. Among these, the movie domain is dominating, and a large number of papers rely on one of the MovieLens datasets (Harper and Konstan 2015). A set of works tackles the issue of popularity bias in the context of e-commerce (Gupta et al. 2019; Wang and Wang 2022; Luo and Wu 2023; Guo et al. 2023; Liu et al. 2023a), and a few works concentrate on the tourism-related problem of POI recommendation (Banerjee et al. 2020; Sánchez and Bellogín 2021; Rahmani et al. 2022a, b). For a number of other application domains, only one or a few research works were identified. We categorized them as "other" application domains, which for example include fashion (Lee et al. 2021), scientific articles (Yang et al. 2018b), jokes (Chong and Abeliuk 2019), or games (Jadidinejad et al. 2019).

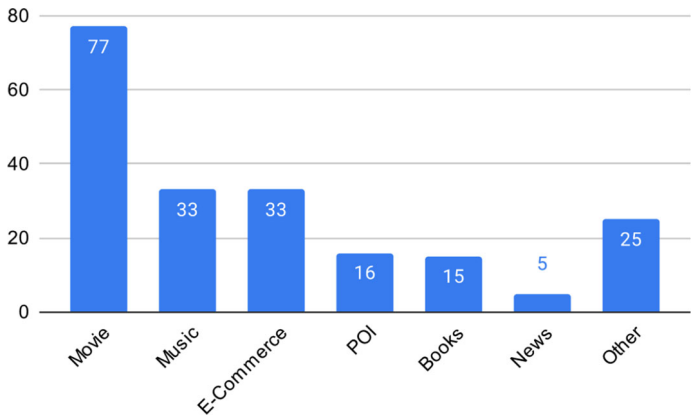

**Fig. 6** Application domains

During the investigation of the papers considered in this survey we noticed that a number of papers[4] provide no specific argumentation why popularity bias can be harmful *in the given application domain* or why a specific dataset is used for the evaluation. In other cases, authors argue that popularity bias might be especially harmful in certain domains, while presenting their work based on data from domains for which it may not be immediately clear what significant harms may emerge from popularity bias, e.g., movie recommendations. According to our discussion above in Sect. 4.1, we often found that the research motivation is given mostly in broad terms (e.g., that the recommendations contain too many popular items or that the "rich get richer").

### 4.4 Types of contributions

Finally, to better understand the landscape of existing research, we characterized the identified papers in terms of their contribution. We identified three main classes of such contributions based on the analysis of the main novel aspects of the papers:

– Papers that *analyze* or *quantify* potentially existing biases;
– Papers that make technical proposals to *mitigate* existing biases;
– Papers that try to *utilize* popularity information to improve recommendations.

---

[4] We deliberately refrain from singling out individual papers here. The list of papers considered in this survey can be inspected online.

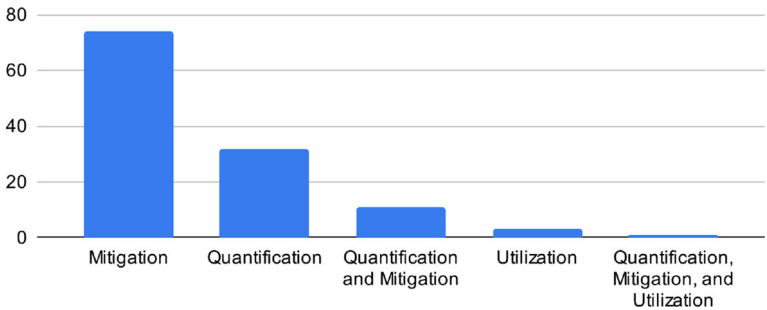

**Fig. 7** Types of research contributions

Figure 7 provides the statistics of the studied papers in terms of this categorization. The detailed categorization of the analyzed works in terms of the contribution can be found in Table 1. We note that one paper can fall into more than one category. Not surprisingly, since we focus on papers in the area of computer science, the majority of papers propose a technical approach to *mitigate* some potential harms of popularity bias. A smaller number of works aim to mainly *quantify and analyze* existing biases in datasets and/or propose computational metrics to assess the extent of the bias. Finally, a limited number of works try to *utilize* information about the general popularity of an item for improved recommendations. We will review selected works in each category next.

## 5 Technical approaches to deal with popularity bias

In this section, we discuss a number of selected approaches to bias quantification, mitigation, and utilization in more depth.

### 5.1 Bias quantification approaches

Papers in this category mainly aim to understand the extent and severity of a possible existing popularity bias and how such bias may impact users.

Before we review existing works that quantify popularity bias for different purposes, we note that any quantification approach—as well as mitigation techniques which we discuss later—requires the definition of appropriate metrics. We will review a multitude of metrics later in Sect. 6. According to our notion of popularity bias from above, these metrics primarily quantify *popularity properties of the recommendations* and not, for example, of the underlying data. However, properties of the underlying data are central in many works, for example when it comes to deciding if an item is considered popular or not. A common strategy in the literature is to categorize items as being popular (short head) or unpopular (long tail), occasionally with an additional separation of the long tail into a middle part and distant tail, see Abdollahpouri et al. (2019b), Borges and Stefanidis (2020). Commonly, this separation is based on the number of observed interactions for each item in the dataset. Yalcin (2021), in contrast, uses a definition where *blockbuster* items not only have to have a high number of interactions, but they must have a high average rating as well. In any case, a central question in such approaches is how to define suitable thresholds. In the existing literature, mostly rules of thumb are applied for which no clear reasoning is provided.

**Table 1** Categorization of papers in terms of *types of contribution*

| | |
|---|---|
| Quantification | Yalcin (2021), Zhu et al. (2021a), Abdollahpouri et al. (2019b), Abdollahpouri et al. (2020b), Abdollahpouri et al. (2021), Neophytou et al. (2022), Zhu et al. (2021a), Elahi et al. (2021b), Lacic et al. (2021), Lesota et al. (2021), Ekstrand et al. (2018), Rahmani et al. (2022b), Zhang et al. (2021), Deldjoo et al. (2021), Channamsetty and Ekstrand (2017), Celma and Cano (2008), Borges and Stefanidis (2020), Borges and Stefanidis (2021), Mena-Maldonado et al. (2021), Heuer et al. (2021), Sánchez and Bellogín (2021), Kowald and Lacic (2022), Naghiaei et al. (2022), Kowald et al. (2020), Vall et al. (2019), Guíñez et al. (2021), Banerjee et al. (2020), Chong and Abeliuk (2019), Mansoury et al. (2020b), Channamsetty and Ekstrand (2017), Yang et al. (2018b), Ferwerda et al. (2023), Li et al. (2023), Lesota et al. (2023), Ohsaka and Togashi (2023), Yu et al. (2022), Kowald et al. (2023), Guo et al. (2023), Chen et al. (2024), Nguyen et al. (2023), Tacli et al. (2022), Lesota et al. (2022) |
| Mitigation | Cremonesi et al. (2014), Jadidinejad et al. (2019), Seki and Maehara (2020), Boratto et al. (2021), He et al. (2022), Abdollahpouri et al. (2017a), Zhu et al. (2021a), Adomavicius and Kwon (2011), Zhu et al. (2021b), Yalcin and Bilge (2021), Wei et al. (2021), Zhang et al. (2021), Seymen et al. (2021), Zhao et al. (2013), Oh et al. (2011), Bedi et al. (2014), Borges and Stefanidis (2021), Gupta et al. (2019), Gangwar and Jain (2021), Huang et al. (2022), Wan et al. (2022), Zheng et al. (2021), Parapar and Radlinski (2021), Zhou et al. (2020), Wu et al. (2019), Yang et al. (2018a), Wang and Wang (2022), Sandholm and Ung (2011), Chen et al. (2014), Hansen et al. (2021), Gharahighehi et al. (2021), Boratto et al. (2021), Rahmani et al. (2022a), Lee and Lee (2015), Shrivastava et al. (2022), Hou et al. (2018), Sharma and Bedi (2018), Sun and Xu (2019), Kelen and Benczúr (2021), Cagali et al. (2021), Schnabel et al. (2016), Saito (2020), Lee et al. (2021), Yin et al. (2012), Li et al. (2021), Kamishima et al. (2014), Abdollahpouri et al. (2021), Cremonesi et al. (2014), Abdollahpouri and Burke (2019), Zhu et al. (2021b), Abdollahpouri et al. (2019a), Mansoury et al. (2020a), Wang and Wang (2022), Wang et al. (2022a), Eskandanian and Mobasher (2020), Yalcin and Bilge (2022), Zanon et al. (2022), Dong et al. (2019), Klimashevskaia et al. (2022), Yalcin (2022), Klimashevskaia et al. (2023a), Sultan et al. (2022), Nguyen et al. (2023), Kim et al. (2023), Rhee et al. (2022), Lin et al. (2022), Liu et al. (2023b), Li et al. (2023), Yang et al. (2023b), Chen et al. (2023), Luo and Wu (2023), Zhang and Shen (2023), Yang et al. (2023a), Klimashevskaia et al. (2023b), Guo et al. (2023), Liu et al. (2023a), Gupta et al. (2023), Chen et al. (2024), Wang (2023), Jia et al. (2023), Ren et al. (2023), Shi et al. (2024), Zheng et al. (2023), Kou et al. (2022), Liu et al. (2022), Ihemelandu and Ekstrand (2023) |
| Utilization | Qi et al. (2021), Zhang et al. (2021), Park and Tuzhilin (2008), Zhao et al. (2022), Zhang et al. (2022) |

An additional approach to quantify popularity-based phenomena is proposed in Celma and Cano (2008) and Celma and Herrera (2008). In their work in the music domain, the authors not only use playcounts as popularity indicators but also rely on metrics from *complex network analysis* to model the *connectedness* of items based on their similarity. This, for example, allows them to analyze if the most popular items are mainly connected to other popular items as well, and to assess the chances if an item being exposed and discovered through recommendations.

*Quantifying Effects on Users.* One common goal in the literature in this area is to mainly quantify the extent of the popularity bias, and in many cases, these observations are then contrasted with other metrics such as accuracy. In such works, often a variety of algorithms from different families, e.g., collaborative and content-based, are compared on different datasets, see, e.g., Channamsetty and Ekstrand (2017), Chong and Abeliuk (2019), Vall et al. (2019). The analysis in Jannach et al. (2015) furthermore shows that even algorithms from the same family, in that case collaborative filtering, can exhibit quite different tendencies to recommend popular items.

While these works usually measure popularity bias across the entire user base, there are a number of works that consider certain subgroups individually. Some works identify such subgroups based on demographics, e.g., based on age and gender (Ekstrand et al. 2018; Lesota et al. 2021; Neophytou et al. 2022) or language (Elahi et al. 2021b). In these works, the goal often is to assess to what extent popularity bias affects the utility of the provided recommendations for different subgroups. The findings in Ekstrand et al. (2018), for example, suggest that there is a non-trivial, and possibly detrimental, interaction of demographics with popularity bias. Elahi et al. (2021b), on the other hand, performed a comprehensive study on popularity bias and investigated, among other aspects, if the strength of bias effects is related to the user's language. Their analyses based on Twitter data indeed indicate that language may play a role and that some effects are more pronounced for English than for other languages. Finally, Sánchez and Bellogín (2021) assessed the effect of popularity bias in Point-of-Interest recommendation on two different user segments: tourists and locals. Their analyses indicate that the utility of the recommendations declines for the latter group of users.

An alternative to segmenting users based on their properties or demographics is to group them based on their preferences or behavior. Some users may, for example, have a tendency to mostly watch mainstream movies, whereas others may exhibit a preference for niche movies. Recommender systems can analyze the user profiles in this respect and categorize them according to their *popularity tendency* or *mainstreamness*. Taking such user-individual preferences into account is central to *calibration* approaches, see Oh et al. (2011), Steck (2018), Abdollahpouri et al. (2020b). One important question in such research works is if certain groups of users—in particular niche item lovers—receive less utility from the recommendations than others. In a number of works such phenomena are seen as a form of potential discrimination, leading to questions of fairness in recommender systems and its relationship to popularity bias (Abdollahpouri et al. 2019b; Borges and Stefanidis 2020; Kowald and Lacic 2022; Kowald et al. 2020; Naghiaei et al. 2022; Rahmani et al. 2022b).

*Understanding Longitudinal Effects.* Most of the works discussed so far adopt a static perspective, e.g., by assessing the popularity bias of a given algorithm at a certain point in time. One main problem of popularity bias however lies in the feedback loop that it can create, which cannot be directly assessed with such forms of "one-shot" evaluations. A number of research works therefore try to study longitudinal effects of biased recommendations. A common way to address such issues in the literature is to rely on a simulation approach. In Jannach et al. (2015), for example, it is assumed that users of a recommender system accept some item suggestions with a certain probability, and that they then provide feedback to the system in terms of ratings, which is fed back into the training data and recommendation model, see also Adomavicius et al. (2021). The results of the simulation indicate that different algorithms can either reinforce or reduce popularity bias over time. Later simulation approaches following similar ideas are presented in Chong and Abeliuk (2019) and in Mansoury et al. (2020b).

A quite different approach to study popularity bias over time was followed in Heuer et al. (2021). In their work, the authors use an auditing approach to assess bias amplification effects on YouTube. Technically, they simulate the user experience with bots that perform random walks over recommended videos on a certain topic. One part of their findings suggests that *"YouTube is recommending increasingly popular but topically unrelated videos"*. Overall, the work is one of the few works in which popularity bias is studied "in-the-wild".

*Popularity Aspects as Performance Predictors.* Finally, some researchers quantify popularity bias in a given dataset with the goal of predicting the performance of different recommendation algorithms. The popularity distribution of the items was for example examined in Deldjoo et al. (2021) as one of several data characteristics that can impact the accuracy of the model. The experimental analysis indeed indicated that the various metrics that capture the characteristics of the popularity distribution can be helpful to contribute to accurate predictions. This seems, in particular, true for algorithms that are known to have a certain tendency towards popular items such as Bayesian Personalized Ranking (Rendle et al. 2009). A related analysis on the impact of dataset characteristics on algorithm performance can be found in Adomavicius and Zhang (2012), where the distribution of the ratings was used as a predictor in the form of the Gini index.

### 5.2 Bias mitigation approaches

Here, we will first categorize existing works based on the processing stage in which the bias is mitigated. Next, we will review a number of technical approaches in more depth.

#### 5.2.1 Categorization per processing stage

As indicated in Fig. 7, the majority of published papers are devoted to the problem of *mitigating* existing biases. In this section, we will discuss these technical approaches in more depth. Inspired by the work by Adomavicius and Tuzhilin (2015) on context-aware recommender systems, we categorize existing approaches according to the *processing stage* in which a mitigation strategy is implemented within a recommendation algorithm.

We differentiate between *pre-processing*, *in-processing*, and *post-processing* approaches. Roughly speaking, pre-processing means that the underlying dataset is adapted or filtered in a way before the learning phase. In a simplistic approach, one could, for example, disallow certain very popular items to be recommended in advance. In in-processing approaches, in contrast, the mitigation technique is part of the learning process, e.g., by considering item popularity in the loss function. In post-processing approaches, finally, often an accuracy-optimized list is adapted to account for biases, e.g., by re-ranking the items in a way that less popular items are brought to the front of the list.

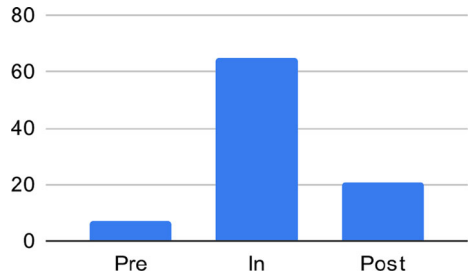

**Fig. 8** Categorization of approaches by processing stage

Figure 8 shows the distribution of papers that propose mitigation strategies according to the processing stage. The detailed categorization per paper can be found in Table 2. We note here that the assignment of individual papers to certain categories in certain cases is subject to a certain level of interpretation. This is in particular the case when it comes to distinguishing between in-processing and post-processing approaches. Even more, when taking an entirely system output-oriented perspective, one could consider almost every approach as being of type *in-processing*. Our chosen categorization of individual papers can be found in Table 2, where one paper can also be assigned to more than one category. In the following, we review selected approaches from the different categories.

### 5.2.2 Pre-processing approaches

Pre-processing approaches to bias mitigation are the least common techniques in our survey. Plus, in many cases, such pre-processing techniques are complemented with additional in-process mitigation steps. Therefore, distinguishing between pre-processing and in-processing techniques often leaves some room for interpretation.

However, at least some approaches—in particular those that apply certain forms of dataset manipulation before model training—can be clearly considered to be pre-processing. Typical pre-processing steps include data sampling, item exclusion, or specific forms of creating positive–negative sample pairs for learning. In Cremonesi et al. (2014), for example, the authors describe an experiment in which the "short head" of highly popular items is removed from the catalogue. The goal of their work was to investigate through a user study how the user experience and the perceived utility of a recommender system changes when those highly-popular items are not recommended.

A lighter form of data sampling was applied in Seki and Maehara (2020). Here, the goal of the pre-processing step is to create a balanced dataset in order to mitigate different fairness issues, with popularity bias being one of them. Ultimately, through the balancing process, the authors aim to create fairer models. However, it has to be noted that such data sampling and balancing must be done with care, in particular to ensure that the remaining data are still representative.

**Table 2** Categorization of papers per processing stage

| Mitigation | Pre-Processing | Boratto et al. (2021), Cremonesi et al. (2014), He et al. (2022), Jadidine-jad et al. (2019), Seki and Mae-hara (2020), Ihemelandu and Ekstrand (2023) |
| | In-Processing | Abdollahpouri et al. (2017a), Zhu et al. (2021a), Adomavicius and Kwon (2011), Zhu et al. (2021b), Yalcin and Bilge (2021), Wei et al. (2021), Zhang et al. (2021), Seymen et al. (2021), Zhao et al. (2013), Oh et al. (2011), Bedi et al. (2014), Borges and Ste-fanidis (2021), Gupta et al. (2019), Gangwar and Jain (2021), Huang et al. (2022), Wan et al. (2022), Zheng et al. (2021), Parapar and Radlinski (2021), Zhou et al. (2020), Wu et al. (2019), Yang et al. (2018a), Wang and Wang (2022), Sandholm and Ung (2011), Chen et al. (2014), Hansen et al. (2021), Gharahighehi et al. (2021), Boratto et al. (2021), Rahmani et al. (2022a), Lee and Lee (2015), Shrivas-tava et al. (2022), Hou et al. (2018), Sharma and Bedi (2018), Sun and Xu (2019), Kelen and Benczúr (2021), Cagali et al. (2021), Schnabel et al. (2016), Saito (2020), Lee et al. (2021), Yin et al. (2012), Li et al. (2021), Kamishima et al. (2014), Kim et al. (2023), Rhee et al. (2022), Lin et al. (2022), Liu et al. (2023b), Li et al. (2023), Yang et al. (2023b), Chen et al. (2023), Luo and Wu (2023), Zhang and Shen (2023), Yang et al. (2023a), Klimashevskaia et al. (2023b), Guo et al. (2023), Liu et al. (2023a), Gupta et al. (2023), Chen et al. (2024), Wang (2023), Jia et al. (2023), Ren et al. (2022), Shi et al. (2024), Zheng et al. (2023), Kou et al. (2022), Liu et al. (2022) |
| | Post-Processing | Abdollahpouri et al. (2021), Cre-monesi et al. (2014), Abdollahpouri and Burke (2019), Zhu et al. (2021b), Abdollahpouri et al. (2019a), Man-soury et al. (2020a), Wang and Wang (2022), Wang et al. (2022a), Eskan-danian and Mobasher (2020), Yalcin and Bilge (2022), Zanon et al. (2022), Dong et al. (2019), Klimashevskaia et al. (2022), Yalcin (2022), Klima-shevskaia et al. (2023a), Sultan et al. (2022), Nguyen et al. (2023) |

Instead of reducing the data with sampling, some authors propose to *augment* the existing data through a pre-processing step, extending the original dataset with additional information beyond interaction data. Such an augmentation could consist of incorporating certain types of item metadata for multi-modality (Cagali et al. 2021); adding information about the users from external sources like their social connections (Li et al. 2021), or combining implicit and explicit feedback as done, e.g., in Jadidinejad et al. (2019). In this latter work, considering rating data is assumed to be useful to (a) more often recommend high-quality items regardless of their (current) popularity and to (b) better leverage existing user feedback during model training.

Park and Tuzhilin (2008) proposed a rather different approach that focuses on enriching the data associated with tail items. According to this approach, the item catalog is partitioned into the head items and the tail items, and the recommendations for each partition are made differently. For the tail items, the recommendation is made by following a method called *Total Clustering (TC)*, which applies clustering techniques to the tail items and then generates recommendations based on the ratings within each cluster. For the head items, on the other hand, a method called *Each Item (EI)* is followed, which applies no clustering and generates recommendations solely based on the ratings of individual items.

An example of a bias mitigation approach that—also according to the authors— has both a pre-processing and an in-processing element is described in Boratto et al. (2021). In what is considered the pre-processing operation, the authors propose specific sampling strategies both for point-wise and pair-wise optimization settings. In the case of pair-wise sampling, for example, the creation of item pairs for learning is not done randomly but depending on item popularity. A similar approach was proposed earlier in Jannach et al. (2015) for the Bayesian Personalized Ranking method.

### 5.2.3 In-process/modeling approaches

In-process approaches are the most common techniques for popularity bias mitigation in the literature. While a variety of in-process techniques were proposed for different application domains and scenarios, they share a common principle, i.e., intervening in the recommendation model to minimize the influence of popular items so that bias is expectedly propagated less through the recommendations. In the following, we discuss the most common families of in-process bias mitigation approaches.

*Regularization-based Approaches* are a prominent group of methods for controlling the influence of popularity (Abdollahpouri et al. 2017a; Kiswanto et al. 2018; Boratto et al. 2021; Zhu et al. 2021b; Kamishima et al. 2014; Seymen et al. 2021). Regularization typically entails adding a term to the optimization objective that lowers the effect of item popularity on the predicted item score. During the learning process, the regularization term thus penalizes the recommendation of popular items and/or helps to promote the less popular items. A specific weight factor (or: coefficient) is often added to the term to adjust the strength of the regularization and thereby balance the competing goals of accuracy and popularity bias.

In an early work in that area, Kamishima et al. (2013), for example, proposed to use a specific regularization term in the optimization objective to build "information-neutral" recommendation systems. Information neutrality means that certain predefined features, as specified by the users, do not influence the recommendation outputs to a significant extent. This idea, which was initially developed in the context of the filter bubble phenomenon, was subsequently applied to the problem of popularity bias in Kamishima et al. (2014), where the goal correspondingly is to end up with a popularity-neutral recommender system.

Later on, inspired by earlier work on dealing with accuracy-*diversity* trade-offs, Abdollahpouri et al. (2017a) proposed to balance popularity and accuracy through a regularization term that penalizes the recommendation of popular items in learning-to-rank approaches. We note that popularity bias is considered a fairness issue in their work, and that considering less popular items in the recommendations is mostly equated with increased fairness.

Boratto et al. (2021) and Zhu et al. (2021b) recently proposed "correlation-based" regularization approaches for combining the predicted scores and item popularity values. In these approaches, the influence of popularity is reduced by applying a penalty when the relevance score for an item is predicted to be high primarily *due to its popularity*. Technically, these approaches build on an idea that was proposed earlier in Beutel et al. (2019) for increasing the fairness in recommender systems.

*Constraint-based Approaches* in general take into account a set of rules (constraints) in order to limit the space of solutions and guide the learning process of a model toward a more efficient and accurate result. As an example, Wang and Wang (2022) introduced the concept of $(\alpha, \beta)$-fairness, which posits that "*similar items should receive similar coverage in the recommendations*". The goal of the approach, where the parameters $\alpha$ and $\beta$ determine item similarity and coverage similarity, is to equalize the exposure level among similar items. By embedding this constraint into a stochastic policy of a deep learning recommendation model, the popularity bias can be reduced.

Another notable constraint-based approach was proposed in Seymen et al. (2021), where a technique to combine constraints and optimization tasks was adopted in a recommendation framework. In particular, the proposed technique extends the optimization objective for recommendation with a set of decision variables that define various constraints, e.g., upper and lower bounds, auxiliary variables, and weighted sums to adjust and control various features of the recommendations. For example, the general popularity of the recommendation can be controlled with an upper bound enforcing the recommendations to contain less popular items. The framework is versatile in the sense that various types of constraints can be easily incorporated. In their paper, the authors used the framework to address different problems and tasks, including provider fairness, popularity bias, and diversification.

*Re-Weighting Approaches* control the effect of popularity by adjusting the weights in the recommendation model in certain ways (Gharahighehi et al. 2021; Steck 2011; Zhao et al. 2013; Gangwar and Jain 2021; Bedi et al. 2014). One early re-weighting approach was proposed by Steck (2011). In this work, the trade-off between recommending long-tail items and accuracy is examined. To address this issue, the author suggests a new metric called "popularity-stratified recall", which combines the two objectives in a single performance measure in a way that recommendations from the long tail are considered to be more valuable. During training, one can then either decrease the weights of the (many) observed ratings for the popular items or increase the weights of the *imputed* (missing) ratings in the ranking process, see also Steck (2010). A notable aspect of the work in Steck (2011) is that it reports the outcomes of an initial study with users. The study indicated that at least for this particular study setup, the users appreciated only a light bias towards less popular items.

Down-weighting the popular items was also proposed by Zhao et al. (2013), where the authors propose a weight adjustment mechanism that can leverage a number of factors reflective of the collected user data, e.g., the opinions of the users, co-rating information, and the values of the ratings provided by users. An example of a work that uses the opposite approach of up-weighting long tail items can be found in Gangwar and Jain (2021), where a boosting algorithm inspired by Schapire (1999) is used to adjust the weights to boost the exposure of the less popular items.

We note here that many re-weighting works discussed above adopt a *static* approach to assess the effects of popularity bias, Zhu et al. (2021a) adopt a longitudinal perspective on the development of popularity bias over time, see also Jannach et al. (2015), Ferraro et al. (2020). The rationale behind the work was that in real recommender systems the users repeatedly receive recommendations that are not necessarily interesting to them and hence have never been consumed by them. Such recommendations represent a false positive error and hence can be used as a source of negative feedback data. As a result, the probability of the user liking such a recommendation decreases with every new recommendation presented to the user. Following this idea and corresponding simulation results, the authors propose to gradually increase the debiasing strength of an underlying re-weighting (or: re-scaling) over time through a dynamically changing hyperparameter.

What can be considered a special case of re-weighting are methods based on Inverse Propensity Scoring (IPS) (Schnabel et al. 2016; Huang et al. 2022; Lee et al. 2021). The concept of inverse propensity has been adopted from statistics and utilized in several prior works to reduce the influence of popularity. In the context of recommender systems, the propensity score can be defined as the probability that a user will find a particular item interesting, hence, like it, based on the observed characteristics and behavior of the user. The propensity score is often based on the popularity of the items (Lee et al. 2021), as users are more likely to interact with popular items in general. Applying the inverse of this score as a penalty then helps to avoid that the recommendation model overestimates the relevance of the observations for generally popular items.

Schnabel et al. (2016) are among the first who considered propensity scores to increase exposure for certain groups of items and hence mitigate selection bias. This is a phenomenon that is commonly believed to be tightly connected with popularity bias. Additionally, this work utilizes causal inference and counterfactual reasoning for unbiased recommendation quality estimation. Huang et al. (2022) later described a related approach which additionally considers the dynamic aspect of propensity scoring. The authors argue that recommendation algorithms should account for user preference changes over time. Both of the previously discussed works are based on explicit user ratings. Lee et al. (2021), in contrast, base their propensity scoring approach based on implicit feedback (click data). This approach extends the commonly adopted positive propensities by considering negative propensities from missing data. The authors suggest that the meaning of the missing feedback is initially ambiguous—it is unclear whether it is negative feedback or just a yet unseen item. Thus, learning to estimate true positive and true negative preferences from both clicked and missing data in an unbiased way has the potential to improve the accuracy of recommendations significantly.

Unbiased, and thus more accurate, recommendations are also the focus in Wan et al. (2022). In this work, Wan et al. propose a modified loss function, named "cross pairwise" loss. The authors argue that cross-pairwise loss is less prone to bias than pairwise or pointwise loss approaches since it can better optimize the predicted scores towards true relevancy scores. Furthermore, it is assumed that the proposed technique can overcome some of the limitations of IPS-based methods, namely, eliminating the need to define propensities in order to describe the exposure mechanism for the recommendation model. Generally, a limitation of propensity-based techniques is that the actual values of the propensities are initially unknown and hence need to be approximated. This makes these techniques becoming sensitive to the choice of the propensity estimator and hence suffer from potential bias in estimation, estimation errors, and propensity misspecification (Yang et al. 2018b). Saito (2020) therefore suggested a propensity-independent loss function to address these potential limitations of IPS-based methods.

*Graph-based Similarity Adjustment* is used to control the influence of popularity bias in graph-based recommender systems, e.g., in Hou et al. (2018), Chen et al. (2014), by "correcting" the way item or user similarity is defined. Chen et al. in Chen et al. (2014) suggest an alternative to cosine similarity, which is typically used for graph-based collaborative filtering algorithms. The new similarity measure accounts for two important factors: user taste, which is represented by the user node degree, and item popularity, which is measured by item node degree. Including these two terms into a new similarity measure and controlling these terms with adjustable coefficients allows to define how strongly these factors influence the predicted score. This, in return, helps mitigating popularity bias and reducing the power of popularity. Another work using item node degree is Hou et al. (2018), which also proposes using a novel similarity measure. The authors name it "balanced similarity index" and state that their approach is able to put more focus on items which are neither extremely popular nor unpopular. Both mentioned approaches use a coefficient to control the debiasing strength, which has to be fine-tuned to find the best trade-off between recommendation accuracy and popularity bias mitigation.

*Integration of Side Information* in the recommendation process is another approach that is utilized to address problems of popularity bias. The strategies that follow this approach may have originally been devised to focus on different objectives, e.g., user preference satisfaction, recommendation novelty, and diversity. However, they exhibit effectiveness in mitigating the popularity bias as well. The rationale behind the consideration of side information is that the lack of sufficient interaction data, e.g., in collaborative filtering techniques, can affect the unpopular (long tail) items more and cause them to be underexposed in the recommendations. For example, it has been shown that popular items tend to have more *neighborhood* relationships than unpopular items (Hou et al. 2018). This can result in lower degrees of similarities among unpopular items leading them to be overlooked in the generation of recommendation by neighborhood-based collaborative filtering. As a matter of fact, pure collaborative filtering techniques are generally considered to be more prone to reinforce popularity bias due to their sole reliance on user interaction data (Jannach et al. 2015). Extending these techniques by incorporating additional features in the recommendation process may help with problem and compensate for the missing data. This can particularly be useful when computing item-based or user-based relationships, by incorporating such additional information, e.g., social connections of users or description of item products. Thus, incorporating *side information* may help in better balancing the inclusion of both popular and unpopular items in the recommendation and mitigating popularity bias.

We note that we use the term 'side information' both for structured item meta-data, as well as for textual information related to the items, such as content descriptions or reviews. Since textual side information is commonly processed with specific natural language processing (NLP) algorithms, we will discuss these approaches separately later.

In an earlier work in that direction, Sandholm and Ung (2011) propose a model to generate real-time location-aware recommendations by incorporating item popularity. The approach forces the recommender to put more emphasis on the location-based relevance of an item, instead of promoting something highly popular but essentially irrelevant due to the user's current location. Similarly, Rahmani et al. (2022a) utilize a set of contextual features for Point-of-Interest (POI) recommendation. Their approach incorporates not only geographical, but also social and temporal context information, combining them with context fusion. The authors then demonstrate how contextualized POI recommendations are less vulnerable to popularity bias than classic collaborative filtering approaches, even when no explicit mitigation approach is applied. Another approach proposed by Sun and Xu (2019) is a topic-based model enriched by incorporating social relations of users. A main assumption of this work is that modeling social relations can assist the recommender system in dealing with the lack of user interaction data for unpopular items, which in turn helps alleviate popularity bias.

Some of the works that rely on side information are not primarily focusing on lowering popularity bias. Instead, they focus on improving novelty or diversity. However, these aspects are often measured in terms of metrics that are based on item popularity statistics. Examples of such research works are Cagali et al. (2021), Yang et al. (2018a), Hansen et al. (2021), which propose multi-modal frameworks to enrich the recommendation model with various types of information for improving the recommendation quality. In Hansen et al. (2021), a model is designed to summarize sessions and create a dynamic user representation based on session interaction sequences. Building on that, the authors propose to combine multiple objectives based on diversity and relevance, using different user and item related features in the music domain. In Cagali et al. (2021) a TV-domain recommendation model is put forward based on different sources of data, including textual, audio-visual, and neural features, together with genre information. Examples of such audio-visual features are chromatic and luminance descriptors of video frames. In Yang et al. (2018a), finally, the authors utilize various types of side information to generate playlist recommendations. The paper incorporates this information in a multi-modal collaborative filtering technique to recommend relevant songs based on a playlist title and content, while keeping the recommendation diverse and novel.

*Natural Language Processing-based Approaches* leverage various kinds of textual information about users or items. One of the most common methods is analyzing textual information contained in user-provided reviews for the items. The approach described in Zhou et al. (2020), for example, relies both on implicit feedback data and review texts. User preference information is first extracted from the user reviews and then fused together with implicit feedback data before the user representation is learned to increase accuracy. Technically, the authors aim to mitigate popularity bias with the help of a two-headed decoder architecture and Noise-Contrastive Estimation (NCE). NCE allows training the model without the explicit assumption that missing interactions indicate a negative preference as done in other models (Wu et al. 2019). This way, missing data for unpopular items will not be automatically dismissed, increasing the accuracy of the recommendations for long-tail items.

Li et al. (2021) employ an autoencoder architecture using text reviews to reconstruct better representations for both users and items. The goal of this work is to optimize the performance of the recommender system for all user groups simultaneously regardless of their "mainstreamness". Shrivastava et al. (2022) propose a similar approach in which opinions and preferences are extracted from user reviews and subsequently combined with rating data. In addition to that, the paper introduces a mechanism to enable the recommendations to optimize multiple objective functions, with the goal of maximizing novelty and serendipity while preserving item relevance. An alternative way of using textual information is proposed in Yin et al. (2012), where topic modeling is applied to classify the items. Technically, Latent Dirichlet Allocation is used to tag items with fine-grained genre-like "topics" to better capture user preferences. This additional meta-data is then used to enrich the recommendation model.

Generally, the described methods are aiding popularity bias mitigation by providing more information extracted from textual data to the recommendation algorithms. This way, they help filling gaps in terms of sparse or missing data for tail items and ultimately enable more accurate representations of user preferences and item characteristics.

*Causal Inference-based Approaches* typically attempt to more deeply investigate the nature of popularity bias itself and what causes it Wei et al. (2021), Zheng et al. (2021), Zhang et al. (2021), He et al. (2022). For example, Wei et al. (2021) model ranking prediction as a cause-and-effect relationship and determine the role of item popularity and user conformity in this relationship. The authors propose to adopt counterfactual inference to mitigate undesired popularity effects. The underlying reasoning for applying a counterfactual approach is that the traditional non-causal learning approach for recommender systems reinforces the observed (factual) user behavior and, as a result, tends to increasingly recommend items because they are popular and not because the item properties match the preferences of a given user. The counterfactual question they therefore seek to answer is what the ranking score would be if the model only focuses on the match between users and items. This information is ultimately used to eliminate or reduce the popularity effect during recommendation.

A similar approach based on counterfactual inference is discussed in He et al. (2022), with a different causality model. Both works introduce a counterfactual world to reduce the influence of popularity on the resulting recommendations.

In a related work, Zheng et al. (2021) adopt causal models to describe how user interactions happen and hence try to attribute them to either user conformity or the true preferences of users. Zhang et al. (2021) also seek to remove the influence of popularity in a causal relationship, while taking into consideration the temporal aspect of recommendation and the fact that item popularity is not a constant. The authors introduce a measure called "popularity drift" to describe the shifting item popularities and predict popularity trends in the future. The authors claim that knowing these trends, a certain part of popularity bias can be actually retained to promote items that have the potential to become popular, but are not yet there and require an exposure boost.

### 5.2.4 Post-processing approaches

Post-processing techniques are quite popular for bias mitigation. The major benefits of post-processing approaches include their typically low cost of implementation, their versatility and their low *intrusiveness*, i.e., post-processing techniques are commonly applied *on top* of an underlying recommendation model. Moreover, some of the existing methods are very general and can be applied in various application domains.

Technically, the main forms of post-processing in the literature are

– re-scaling (score adjustment),
– re-ranking (reordering),
– rank aggregation

All of these methods are commonly based on one given recommendation list ranked by accuracy scores, and they then incorporate additional information in the post-processing phase.
*Re-scaling (or: score adjustment)* works by updating the relevance scores of a given recommendation list to compensate for popularity bias by promoting certain items or penalizing the others. The updated scores are then used to re-order the list of the recommended items. In the case of *re-ranking*, the item order is changed as well, however the original relevance scores are considered less relevant and discarded in some cases. Instead, these approaches often operate solely on the item rank, swapping or exchanging the items to fulfill certain criteria. *Rank aggregation* post-processing involves multiple recommendation lists produced for the same user by different models and is based on fusing these lists with rank aggregation methods. Last but not least, besides the described three methods, a post-filtering technique may simply remove certain (popular) items from a recommendation list.

In *re-scaling* the goal is to boost or penalize certain items in the recommendation list. In the context of popularity consideration this could be seen as a *bias correction* approach. The typical goals, therefore, are to *(a)* include more or less popular items that could be potentially interesting to the user, *(b)* exclude the popular items that the user is not interested in or already knows about anyway, and *(c)* do not include the items that are both unpopular and uninteresting to the user. An example of a recent post-processing approach work can be found in Zhu et al. (2021b), where the authors propose to add a *compensation score* to the predicted preference score in a way to consider the above goals in appropriate ways. We note that in the same work an in-processing approach based on regularization is proposed as well. In another post-processing approach, Zhu et al. (2021a) apply bias correction as well, however with a dynamic perspective, where bias mitigation is applied iteratively and repeatedly over time.

*Re-ranking* appears to be the most common post-processing technique among the reviewed works. Generally, these methods attempt to re-order the items in the recommendation list in such a way that it optimizes for a certain objective metric. For example, the approach described by Abdollahpouri et al. (2021) is targeted towards balancing the relevancy and popularity of items in the list, with a flexible parameter that gives more significance to either of the features. The same objective function has been earlier introduced by Steck (2011) for an in-processing mitigation approach. Klimashevskaia et al. (2022) later on reproduced this approach, demonstrating that even though the method is able to adjust the recommendations to the user popularity preferences, this does not necessarily mitigate platform-wide popularity bias in a significant way. In an earlier work, Abdollahpouri et al. (2019a) proposed an adaptation of the xQuAD query diversification algorithm for popularity bias mitigation. In a related work, the authors also investigated the performance of this method from a longitudinal perspective in Abdollahpouri and Burke (2019).

A number of re-ranking based works connect popularity bias closely to the concept of *novelty*. Both Oh et al. (2011) and Bedi et al. (2014) suggest ways of including more novel and underexposed items in recommendation lists to improve the utility of the recommendations. Other works aim to penalize only specific types of popular items, e.g., "blockbuster" items in Yalcin and Bilge (2022), or implement certain application-specific features or metrics as in Wang et al. (2022a) in the context of crowdworker recommendation.

Finally, some works rely on techniques from graph and network science to rearrange the recommendation lists to achieve certain distribution goals. The visibility of items through bipartite graphs is considered in Mansoury et al. (2020a), and a stable matching algorithm is used in Eskandanian and Mobasher (2020). Both methods represent items and/or users within as nodes of a graph and use this model to investigate and increase the exposure of items in the resulting rearranged recommendation list. Zanon et al. (2022) in contrast, describe a graph-based approach of incorporating additional similarity information for re-ranking.

*Rank Aggregation* works by counteracting the popularity bias introduced during model training by combining it with an alternative ranking. For instance, Dong et al. (2019) suggest combining a given ranking with a reverse recommendation ranking via Two-Way Rank aggregation. Alternatively, item ranking can be also combined with an inverse popularity ranking for a user or a group of users, as proposed in Yalcin and Bilge (2021). A very particular way of relying on multiple ranked lists is proposed in Yalcin (2022). Here, the idea is not to produce multiple lists and to combine them, but to *select* one of the several pre-generated lists based on pre-defined criteria such as preference match, diversity, or popularity distribution.

### 5.3 Bias utilization methods

There are a few works which try to make use of the fact that popular items are by definition liked by many—and are thus also "safe" recommendations, see our discussions above about Netflix adding popularity signal to their video ranker.

Zhao et al. (2022) for example claim that not all item popularity is the same and it may often result from the genuine quality of an item and can thus lead to high-quality recommendations. The authors suggest to leverage this "true quality popularity" and mitigate other effects of popularity bias at the same time, disentangling them from each other. An area where the (recent) popularity of the items can be a highly-important signal is the news recommendation domain. The work in Qi et al. (2021), for example, suggests that using article popularity can actually lead to sufficient topical diversity and coverage. A number of earlier works also demonstrate that considering the recent popularity of an article can be crucial for high recommendation accuracy as well (Hopfgartner et al. 2016; Tavakolifard et al. 2013; Garcin et al. 2013). Similar observations regarding the importance of short-term popularity trends were reported for the e-commerce domain in Jannach et al. (2017).

A very different and malicious way of using the existing popularity bias of certain algorithms is discussed in Zhang et al. (2022). Here the authors describe how popularity bias can be abused in an attack to artificially boost a target item, using the predictable behavior of a biased recommender. This vulnerability can falsely skew the popularity distribution even more, potentially leading to the loss of trustworthiness and hurting provider fairness on the platform as well. Overall, this latter work is a key example that demonstrates the importance of studying, understanding, and being able to control the popularity bias of a recommender system.

## 6 Evaluation approaches

In this section, we review the methodology that is used in the research work on popularity bias. We will first analyze which datasets researchers are using for experiments and evaluation. We will then look closer at which types of studies are performed to evaluate the quality of the recommendations and the effectiveness of popularity bias mitigation approaches.

### 6.1 Datasets

Our analyses in Sect. 4.3 revealed that the literature on popularity bias covers a diverse range of application domains. It also turned out that the potential negative effects of popularity bias *in a given domain* were not always clearly stated in the papers. This phenomenon manifests itself also in the context of the evaluation of newly proposed mitigation approaches. Again, this may point to a certain level of overgeneralization or oversimplification of the problem, where the choice of the evaluation dataset may almost appear arbitrary and where potential idiosyncrasies of a given application are not taken into account.

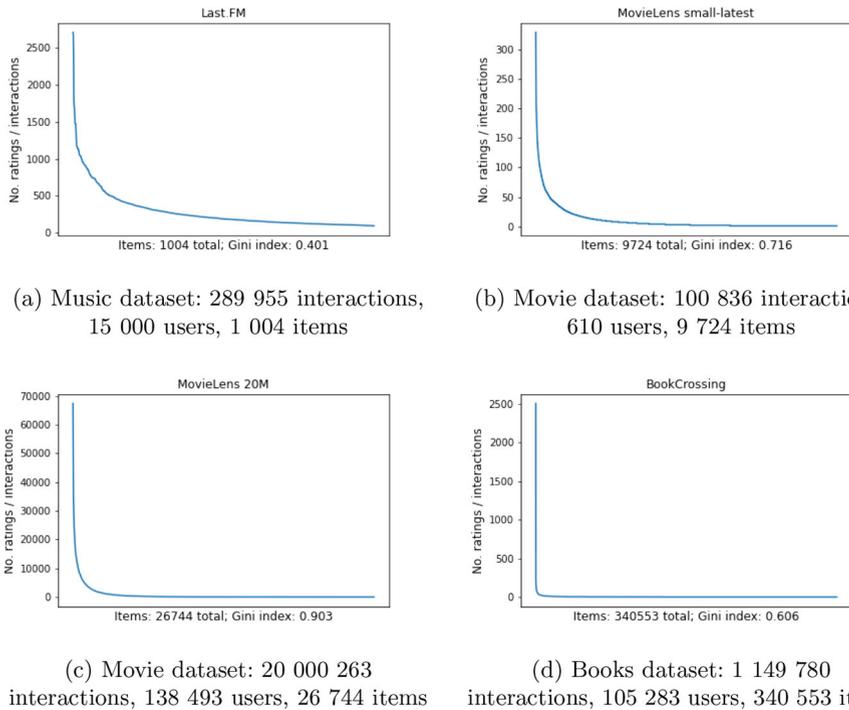

(a) Music dataset: 289 955 interactions, 15 000 users, 1 004 items

(b) Movie dataset: 100 836 interactions, 610 users, 9 724 items

(c) Movie dataset: 20 000 263 interactions, 138 493 users, 26 744 items

(d) Books dataset: 1 149 780 interactions, 105 283 users, 340 553 items

**Fig. 9** Examples of commonly used datasets. The plots show the interaction counts for each item within the dataset on the x-axis, sorted in descending order. The Gini index expresses the inequality of the distribution, with values closer to 1 indicating a high inequality (range: 0–1)

The datasets used for recommender system training and evaluation in the reviewed works all demonstrate skewed popularity distributions to some extent, showing the "long tail curve" (see some examples in Fig. 9). However, they often differ significantly in terms of size, density, and popularity distributions, making it difficult to compare effects and results between datasets. Moreover, researchers sometimes apply additional data pre-processing procedures, which may not always be documented in the papers in detail. Some authors, for example, exclude cold-start items or less active users from the dataset for better training, however, based on different thresholds (Rhee et al. 2022; Lin et al. 2022; Mansoury et al. 2020a; Borges and Stefanidis 2021). These factors may further aggravate the problem of non-comparable evaluation results.

Independent of the different characteristics of the used datasets, an important aspect to question is to what extent these frequently used datasets are truly representative of real-world problems of popularity bias. Datasets like the widely used ones from MovieLens are already pre-filtered and only contain users and items for which a certain number of interactions was recorded. However, in real-world applications, e.g., in e-commerce, only one or a few interactions may be recorded for a large fraction of the users and the items, and some items may have never been purchased during the data collection period (Jannach et al. 2017). Thus, in reality, the popularity distributions might be even more skewed than what we observe in the datasets used in academia.

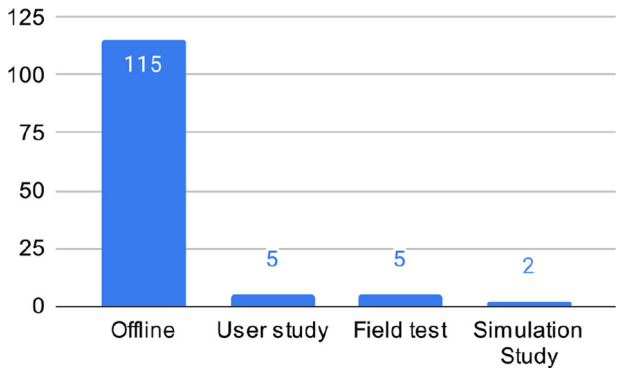

**Fig. 10** Distribution of used evaluation approach in the surveyed papers

Finally, there are certain application domains, which are usually described in the literature as the ones that could potentially experience significant fairness issues due to popularity bias, e.g., job recommendation, healthcare, or banking applications. Unfortunately, public datasets in such domains are very scarce, and it stands to question if the analyses and mitigation techniques that were done in domains like movie recommendation generalize to such critical application areas. We acknowledge how challenging it can be for researchers to obtain or even publish such data. The future availability of data in such domains is however crucial for the development of truly impactful research on fairness-related questions of popularity bias.

## 6.2 Evaluation approaches

Next, we analyze the methodologies researchers rely on when investigating popularity bias mitigation techniques for recommender systems. As done commonly in the literature, we differentiate between offline (data-based) evaluations and studies, user studies (either in the lab or online), and field tests (A/B tests) (Gunawardana et al. 2022). Figure 10 shows that the landscape is very strongly dominated by offline experiments. This is also a general trend to an even larger extent in recommender system research in general (see Jannach et al. (2012) for an earlier survey). Only four works report the outcomes of a user study (Cremonesi et al. 2014; Lee and Lee 2015; Steck 2011; Yin et al. 2012), and a single work was found which examined popularity bias effects in a field test (Lacic et al. 2022). Interestingly, all works that include some form of user study are comparably old and were published in 2015 or earlier. No work considered in our survey relied on alternative qualitative approaches like interviews or observational studies.

We furthermore analyzed if the distribution of applied evaluation approaches may depend on the domain (see Fig. 6) or on the type of research contribution (see Fig. 7). However, we found the same patterns as shown in Fig. 10, i.e., a very strong tendency by researchers to rely on offline evaluations.

In the following, we will discuss selected aspects of both offline studies and studies that involve humans in the loop. We will elaborate on the studies that involve humans in more depth in order to provide examples and raise awareness regarding what kind of research questions can be answered with such studies.

### 6.2.1 Offline evaluation

As noted before, the majority of the studies in this research field have primarily focused on evaluation based on offline experiments. Many studies simply follow a traditional approach adopted from general machine learning research when conducting offline experiments: a pre-collected dataset is split into disjoint subsets for training, validation, and testing. This is frequently done by following common cross-validation methodologies, including k-fold cross-validation, hold-out, and leave-one-out. The split can be performed either randomly (Abdollahpouri et al. 2021; Elahi et al. 2021b; Yin et al. 2012; Zhang et al. 2022; Zhu et al. 2021b; Borges and Stefanidis 2021; Gangwar and Jain 2021; Neophytou et al. 2022; Naghiaei et al. 2022; Kowald and Lacic 2022; Mansoury et al. 2020b; Chong and Abeliuk 2019; Zhao et al. 2013) or chronologically based on the timestamps of the user interactions (Steck 2011; Zhao et al. 2022; Qi et al. 2021; Zhang et al. 2021; Sánchez and Bellogín 2021). This evaluation methodology is also applied using semi-synthetic datasets (Zhu et al. 2021a; Heuer et al. 2021). The quality of recommendation, measured in terms of various evaluation metrics, is then compared before and after bias mitigation strategies are applied to the input or output of the recommender system (i.e., in the *pre-processing* or *post-processing* stage), or directly to the core recommender model (i.e., in the *in-processing* stage).

The impact of popularity bias on different recommender systems and the performance of mitigation strategies can be viewed from *static (one-shot)* and *dynamic (longitudinal)* offline evaluation paradigms. Traditionally, the research community has focused more on the static paradigm. In this case the dataset is split for evaluation randomly and only once, often ignoring the timestamp of the feedback/interactions. Hence, this paradigm reflects the evaluation of a recommender system on an individual *"snapshot"* of the system. Accordingly, the data used for training simulates the knowledge of the recommender about the users given at a certain point in time. The test data respectively simulate the information about users (and their preferences) that is "hidden" from the system at that point in time. The static evaluation paradigm, however, does not reflect temporal changes within the data distributions, and thus the outcomes might be less reliable as a result. Notwithstanding this limitation, this evaluation paradigm may still offer benefits for finding the most suitable design solution for an up-and-running recommender system (e.g., the best-performing algorithm) in certain situations (Zhang et al. 2020).

The dynamic (longitudinal) evaluation paradigm, on the other hand, proposes a radically different perspective that can potentially lead to more trustworthy results. This evaluation paradigm primarily aims at a more continuous and long-term evaluation of a recommender system over a period of time. Hence, the performance of the recommender system is monitored considering the dynamics of the system properties and the data. Examples of the studies employing longitudinal evaluation methodologies are Burke (2010), Ferraro et al. (2020), Jannach et al. (2015), Mansoury et al. (2020b), Heuer et al. (2021), Ohsaka and Togashi (2023), Yu et al. (2022), Shi et al. (2024), Zheng et al. (2023). In this case timestamps are playing a central role in data splitting. Typically, the data is split into $N$ time spans and for every period $n$ the next period $n + 1$ is used as a test set. Afterward, the $n + 1$ subset is appended to the previous training set, the model is retrained on the new extended data and the process is repeated iteratively this way, simulating the temporal evolution of a recommender system (see Fig. 11). It is also possible to simulate user activity by predicting which items from the recommendation for each user will be consumed at every iteration and adding them to an extended train set instead. However, false or inaccurate predictions can lead to errors that might accumulate over time.

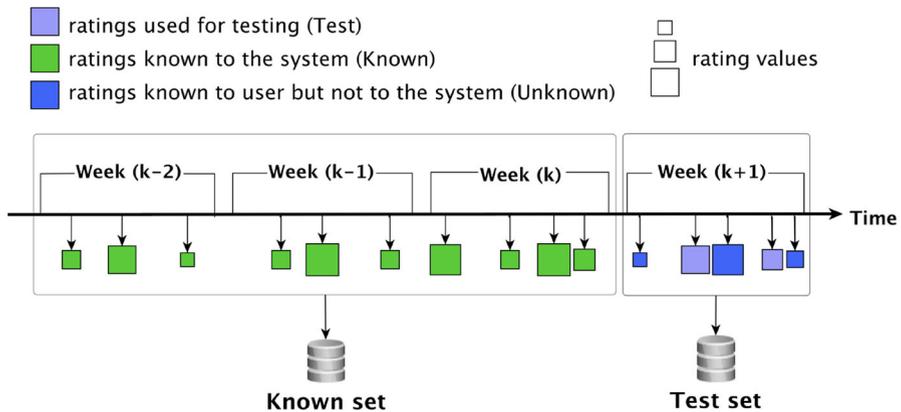

**Fig. 11** An example of week-by-week longitudinal evaluation data split for recommender system training and evaluation; from Elahi (2014)

It is argued that studying the longitudinal evolution of a recommender system provides a better picture of a real-life user experience scenario. In the context of popularity bias mitigation, it can be particularly important to follow the longitudinal evaluation procedure when investigating the "reinforcement effect" of the bias in recommender systems (Ferraro et al. 2020). This will allow obtaining a better reflection on the effectiveness of the mitigation strategies in real-world scenarios, where the behaviors and preferences of the users are constantly changing over time. Shi et al. (2024), for instance, claim that performing repeated iterative simulations make it possible to ensure "long-term user satisfaction" and capture the interactive nature of many recommendation scenarios, where continuous user feedback plays a significant role. Furthermore, Yu et al. (2022) propose a framework to simulate a recommendation environment iteratively with multiple agents as users. Such methods can allow the researchers to get as close as possible to real-life studies without running possibly costly and risky A/B tests.

The differences in the evaluation methodologies make it often difficult to draw a conclusive direct comparison of different bias mitigation strategies. In addition to that, the reported results of the conducted experiments may also differ due to the dissimilarity in the characteristics of the used datasets, the chosen recommender algorithms, and even the choice of the hyper-parameters. For instance, the threshold popularity value used to divide the items into head and tail is an important factor and can substantially impact the outcome of the experiments. Many prior works considered 0.2 as a suitable choice for the threshold (Adomavicius and Kwon 2011; Abdollahpouri et al. 2017a; Kiswanto et al. 2018). Hence, they considered the top 20% of items with the largest number of interactions by users as popular items. At the same time another group of works considered the head part to be represented by the top 10% (Vall et al. 2019) or even 1% of items(Kamishima et al. 2014)—hence they observed experimental outcomes that diverge from the former ones.

Notwithstanding the limitations, offline experiments can offer benefits and be indicative of the general performance of different popularity bias mitigation strategies. Moreover, it is generally agreed that a sound and comprehensive evaluation procedure may include an offline experiment followed up with an online experiment hence applying a three-step methodology (Rashid et al. 2002; Gunawardana and Shani 2015; Carenini et al. 2003; Kluver and Konstan 2014; Zhang and Shen 2023): (i) identifying a set of candidate strategies from the literature and formulating a research hypothesis, (ii) comparing the performance of the candidate strategies through offline experiments based on pre-collected datasets and shortlisting the best-performing strategies, and (iii) conducting follow-up online experiments with real users to verify the impact of the selected strategies.

### 6.2.2 Human-in-the-loop online evaluation

Online evaluation in recommender systems typically involves simulated environments or even real-life settings in which a recommender system is tested by real users. This type of evaluation includes user studies (Cremonesi et al. 2014; Lee and Lee 2015; Steck 2011; Yin et al. 2012; Ferwerda et al. 2023; Lesota et al. 2023) and A/B testing (Lacic et al. 2022; Klimashevskaia et al. 2023a; Zhang and Shen 2023). The former typically requires a prototype, a mock-up recommendation platform or simply a list of provided recommendations that the users are normally required to evaluate and give their opinion on through ratings, feedback or questionnaires. This allows to observe user behavior close to a real-life scenario, without user modeling, predictions or assumptions. Furthermore, user studies allow researchers to gather invaluable information such as user personal opinions and their perception of the recommendation qualities. The downside of online evaluation procedures is often the complexity of the setup. Customarily, a user study platform needs to be deployed and a significant number of testing users need to be incentivized to participate in the study, honestly and diligently following the procedures. These difficulties make online studies more rare and uncommon in recommender system evaluation research—only four works in our literature collection reported results of a user study. Having not many of these works, we can look into more detail about the setups and protocols they are reporting.

The earliest work in our collection that included some form of user study in Steck (2011). While the main focus of the paper is on providing a new "popularity-aware" metric for offline evaluations, the author also reports the initial outcomes of a user study in which 20 subjects participated. The task of the participants was to rank recommended lists with different levels of popularity bias mitigation in terms of recommendation usefulness. Interestingly, it turned out that the already small intervention towards the long tail of the recommendations led to a quickly lowered usefulness perception by the subjects and loss of user trust.

In the work by Yin et al. (2012) extensive offline evaluations are complemented with a user study. In their study, 50 subjects rated movie recommendations that were generated by different algorithms, including ones optimized for long-tail recommendations, from different perspectives such as preference match, novelty, serendipity and overall assessment (quality). The results showed that their proposed method was effective in terms of increasing the novelty and serendipity level of the movies, while still being a good match for the user preferences and leading to recommendations that participants gave a high overall rating.

The work by Cremonesi et al. (2014) is entirely based on a user study. In their case, the authors created a platform simulating hotel recommendation and booking experiences. They conducted an online experiment in which 382 subjects participated, being assigned to one of six experimental groups. Three recommendation algorithms (one of them showing the most popular items) were tested in two scenarios each: (a) recommending accommodations during "low tourist" season, when all hotels are available; (b) recommending in "high tourist" season, when the most popular options are typically already booked and are unavailable. The authors attempted to measure different objective and subjective aspects, with *satisfaction* being the central subjective factors. It turned out that during low season, a non-personalized popular item recommendation strategy was indeed leading to the highest average satisfaction. During high season, however, a hybrid method performed best in this dimension. Overall, it turns out that recommending popular items can be effective in certain cases, and, hence, that popularity bias is not necessarily always bad.

Another online user study was described in Lee and Lee (2015), where the authors built a website that recommended music artists to the users based on their existing profiles on the *last.fm* music service. Recommendations were created through a new algorithm designed for novelty and a baseline MF-based recommender. In total, 44 subjects completed the study in which they were asked to provide feedback on the relevance and *"freshness"* (novelty) of the artists. The obtained results mainly indicated that the new algorithms were effective in increasing novelty at the price of reduced relevance.

Two related user studies by Ferwerda et al. (2023) and Lesota et al. (2023) investigate user perception aspects that can only be measured through questionnaires. Specifically, the authors explore how users perceive recommendation popularity and debiasing in terms of whether these aspects are even noticeable to the end user, and how the perceptions align with calculated popularity bias metrics. Their results indicate that the computational metrics often do not correlate well with what users perceive. Such results emphasize the importance of double-checking how well a metric actually correlates with user perceptions, see also Jesse et al. (2022).

Overall, the user studies discussed so far indicate that there indeed may exist a commonly assumed trade-off between recommendation accuracy and popularity bias mitigation. The studies in Steck (2011) and Cremonesi et al. (2014), however, indicate that focusing more on long tail items can relatively quickly negatively affect the users' perception of the recommendation quality in terms of usefulness or relevance. The drop in mean relevance reported in Lee and Lee (2015) is also not very small, decreasing from 3.8 to 3.3 on a five-point scale. Looking at the scale of the studies, only one Cremonesi et al. (2014) involved a larger sample of participants. In the other cases, mostly a few dozen participants were recruited. Since the user studies in two cases only serve as a complement to offline experiments, few details of the experiments are reported, which can make it difficult to assess to what extent the study might generalize, e.g., to other participant groups.

A/B tests on the other hand, are typically deployed on real-life industry-based platforms using recommender systems. The users on the platform are split into two (rarely more) groups of equal size. One group is a control group receiving ordinary treatment, while the other group would receive recommendation from the algorithm to be tested. In contrast to user studies such an evaluation approach is less invasive and even often performed without the users being aware of it to avoid priming and bias. The main drawback of such evaluation is the possible costs and risks of deploying new approaches on an industry platform, and the opportunity to do so is quite rare in the research community.

In one of the A/B tests among the surveyed papers, Lacic et al. (2022) studied: (a) the effects of the end user devices on item exposure and click-through rates, and (b) the effects of different algorithms on users. A two-week study was conducted on an Austrian newspaper website where a personalized content-based recommender was introduced. From the obtained results the authors conclude that content-based recommendations can reduce the popularity bias for the group of anonymous users over time even during one session. Unfortunately, the study was plagued by two major public events happening during the study period. Also, certain details about the application of the personalized method to anonymous users remained unclear.

Continuing their previous work (Klimashevskaia et al. 2022) on *offline* evaluation of calibrated recommendations, Klimashevskaia et al. (2023a) explored the effects of popularity-based calibration in an online A/B test of the CP re-ranking algorithm (Abdollahpouri et al. 2021). The algorithm was deployed for several months on a real-life movie recommendation and streaming platform to assess to what extent popularity-based re-ranking affects the user experience. Their results showed that the re-ranking approach did not negatively affect recommendation quality (approximated through the click-through-rate), but also enticed users to consume more diverse content.

## 6.3 Evaluation metrics

A range of metrics has been employed by the research community to evaluate the performance of mitigation strategies for popularity bias and to measure the extent of existing bias in the data. These metrics can be grouped in different ways. For instance, from the multi-stakeholder perspective mainly two groups of metrics can be identified, *user-centered* metrics and *item-centered* metrics. While the former group of metrics takes into account the differences among users in terms of their preferences towards popular items, the latter group tends to ignore such differences and concentrates on item qualities instead. It is essential, however, to consider both sides in the evaluation process to assess the effects of the bias and its mitigation in a comprehensive manner (Abdollahpouri et al. 2017b).

**Table 3** Descriptive popularity bias metrics

| Group | Metric Name | Example |
|---|---|---|
| Popularity Bias within the dataset | Gini index | Adomavicius and Kwon (2011) |
| | Popularity skewness/kurtosis | Deldjoo et al. (2021) |
| | Mean/Median Popularity, Popularity Variance | Lesota et al. (2021) |
| | Popularity Bias Evaluation | Deldjoo et al. (2021) |
| | Long Tail Items Evaluation | Deldjoo et al. (2021) |
| | Popularity Drift | Zhang et al. (2021) |
| | Degree of Matthew Effect (DME) | Wang (2023) |
| User Profiling / Categorizing | Shannon entropy | Elahi et al. (2021b) |
| | Personal Popluarity Tendency (PPT) | Oh et al. (2011) |
| | Mainstreamness | Borges and Stefanidis (2020) |
| | Ratio of Popular Item (RPI) | Tacli et al. (2022) |
| | Average Popularity of Rated Items (APRI) | Tacli et al. (2022) |
| | Better-Than-Average propensity (BTA) | Tacli et al. (2022) |
| | Positively-Rated propensity (PR) | Tacli et al. (2022) |

In this work, we, however, adopt an alternative categorization and grouping, based in the main two research goals that we found in the papers that we analyzed for our survey:

– Some metrics are purely descriptive and are commonly utilized for bias characterization and item/user profiling. This includes metrics describing popularity distributions within datasets, such as Popularity skewness, or metrics that describe user profiles like Personal Popularity Tendency or Mainstreamness, see Table 3.

– Other metrics are instead predominantly used as objectives for the popularity bias mitigation process. Item-related examples of such metrics include Catalog Coverage, Average Recommendation Popularity or Item Statistical parity, see Table 4. Metrics like Miscalibration or User Popularity Deviation, on the other hand, can serve as user-centered optimization goals for bias mitigation.

We note that the descriptive metrics can be calculated based solely on the given interaction data. The metrics that are used for steering the mitigation process commonly require a recommendation model or a simulation of a recommendation process to be assessed.

Table 3 shows a list of *descriptive metrics* that we found through our literature survey. The entries in the table are organized in two subcategories for the item and user perspective, respectively.

In Table 4 we list the metrics that are used as *optimization targets* for bias mitigation. The metrics in this table are organized in four subcategories. Metrics in the subcategory "Recommendation Popularity Level" measure how popular the generated recommendations are. Metrics in the category "Catalogue Coverage and Distribution in Recommendations" describe the fraction of categories that actually appear in the recommendations (coverage) and how often they appear (distribution). Metrics in the group "Recommendation Personalization" determine how close the item popularity distribution in the recommendations is to the user preference. Finally, metrics in the last category, "Tail Item Prediction" assess how much the accuracy of the recommendations is affected by item popularity or unpopularity.

Both in Table 3 and in Table 4 we provide example papers in which the metric is used. The technical descriptions of each metric can be found in the referenced literature. We note that some metrics can appear in both tables. The Gini index, for example, can be used to *quantify* the existing unevenness of the popularity distribution in a given dataset. It can, however, also serve as a measure to determine the unevenness of the popularity distribution *of the recommendations* provided by the system.

Overall, we observe that a rich variety of metrics and variations thereof is used in the literature, which makes it often difficult to compare the outcomes of different studies. We note that the variety of metrics is actually even higher as indicated in the tables, as we can find different implementations for some of the metrics as well. For example, the popularity of the items is often measured by the number of interactions recorded for each item in the dataset. In some cases, however, these interaction counts are normalized, whereas in others they are not. Furthermore, sometimes, special metrics like the *Blockbuster Score* (Yalcin 2021) are used as well to assess the popularity of an individual item.

Generally, we find that some metrics are used more frequently than the others. The frequency of different *bias-related* metrics, i.e., popularity and other beyond-accuracy metrics, in the examined papers is shown in Fig. 12a. Figure 12b shows the same chart for the accuracy metrics. For the bias-related metrics, we found that ARP (Average Recommendation Popularity) and the Gini Index are the most frequently used metrics to assess the extent of popularity bias in the data and in the recommendations. Similar to the Gini Index that indicates the distribution of item exposure, Catalogue Coverage is also frequently used for similar purposes, i.e., to estimate how well the catalogue of items is exposed as a whole. These metrics may be considered to be more universal, while many others in some ways depend on how the authors define the bias itself and their mitigation strategy and goals. For instance, APLT (Average Percentage of Long Tail Items) and ACLT (Average Coverage of Long Tail items), both counting long tail items included in recommendation, are frequently used in cases where the authors claim that the focus of mitigation should be on promoting the long tail. Alternatively, if the goal of bias mitigation is generally higher recommendation diversity, then metrics like Aggregate Diversity or Intra-List Diversity are applied.

We furthermore observed that some works use rather case-specific metrics, like the papers describing user studies, which base their measurements on questionnaires and qualitative analyses (Lee and Lee 2015; Cremonesi et al. 2014). Furthermore, works with rather uncommon interpretations or representations of bias also sometimes employ unconventional metrics. Celma and Cano (2008), for example, apply network analysis methods to investigate bias and rely on common metrics from this field of study. Overall, the wide range of used metrics indicates that no commonly-established definition of popularity bias exists in the literature. As shown in Fig. 12a, a relatively large number of bias-related metrics are only used in one single paper, all subsumed in the category 'Other'. Upon closer inspection of these singular metrics, we established that while some of them are truly unique, others can be related and implementing similar ideas with slight differences, like variations on Entropy or Diversity.

**Table 4** Objective-oriented popularity bias metrics

| Group | Subgroup | Metric Name | Example |
|---|---|---|---|
| Recommendation Popularity Level | - | Average Popularity Count, Average Recommendation Popularity (ARP) | Abdollahpouri and Burke (2019) |
| | | Popularity Count (PCount) (Non-normalized ARP) | Borges and Stefanidis (2021) |
| | | Group Average Popularity (GAP) | Kowald et al. (2020) |
| | | Average Percentage of Long Tail Items (APLT) | Abdollahpouri et al. (2019a) |
| | | Popularity Lift | Abdollahpouri et al. (2020b) |
| | | Discounted Cumulative Popularity (DCP) | Borges and Stefanidis (2021) |
| | | Ideal Discounted Cumulative Popularity (IDCP) | Borges and Stefanidis (2021) |
| | | Popularity Bias (POBK) | Borges and Stefanidis (2021) |
| | | Supplier Popularity Deviation (SPD) | Gharahighehi et al. (2021) |
| | | Mean/Median Popularity, Popularity Variance | Lesota et al. (2021) |
| | | PopQ | Rhee et al. (2022) |
| | | Popularity Correlation with Exposure Rate (PER) | Lin et al. (2022) |
| | | Popularity Correlation with Success Rate (PSR) | Lin et al. (2022) |
| Catalogue Coverage and Distribution in Recommendations | Item Distribution and Exposure | Gini index | Adomavicius and Kwon (2011) |
| | | Entropy-Diversity | Adomavicius and Kwon (2011) |
| | | Herfindahl index | Adomavicius and Kwon (2011) |
| | | Coverage Disparity | Wang and Wang (2022) |
| | | fairRate@K | Wang et al. (2022a) |
| | | Equity of Attention for Group Fairness (EAGF) | Gharahighehi et al. (2021) |
| | | Item Statistical Parity (ISP) | Boratto et al. (2021) |
| | | Item Equal Opportunity (IEO) | Boratto et al. (2021) |
| | | Genralized Cross-Entropy (GCE) | Rahmani et al. (2022b) |
| | | Recommendation Ratio | Zhang et al. (2021) |

**Table 4** continued

| Group | Subgroup | Metric Name | Example |
|---|---|---|---|
| | | Popularity-Opportunity | Zhu et al. (2021a) |
| | | (α, β)-fairness | Wang and Wang (2022) |
| | | Bias reduction | Wang and Wang (2022) |
| | | Exposure Bias | Banerjee et al. (2020) |
| | | Popularity Parity | Gupta et al. (2023) |
| | | Ranking-based statistical parity (RSP) | Liu et al. (2023b) |
| | | Ranking-based equal opportunity (REO) | Liu et al. (2023b) |
| | General Catalogue Coverage | Aggregate Diversity / Catalog Coverage | Oh et al. (2011) |
| | | Personalization | Cagali et al. (2021) |
| | Short Head / Long Tail Coverage | Average Coverage of Long Tail items (ACLT/ACT) | Abdollahpouri et al. (2019a) |
| | | Ratio of Popular Items | Yalcin (2021) |
| | | Blockbuster Recommendation Frequency (BRF) | Yalcin (2021) |
| Popularity-Aware Personalization | - | Miscalibration (popularity) | Abdollahpouri et al. (2020b) |
| | | Kullback–Leibler Divergence of Popularity Distributions | Lesota et al. (2021) |
| | | Kendall's Tau of Popularity Distributions | Lesota et al. (2021) |
| | | Mean Absolute Deviation of Ranking Performance | Rahmani et al. (2022b) |
| | | User Popularity Deviation (UPD), temporal version of UPD | Klimashevskaia et al. (2022) |
| | | Temporal version of UPD | Guñez et al. (2021) |
| | | Popularity Correlation with Conversational Utility (PCU) | Lin et al. (2022) |
| | | User-oriented group fairness (UGF) | Liu et al. (2023b) |
| Tail Item Prediction Quality | - | Popularity-Rank Correlation (item- or user-based) | Zhu et al. (2021a) |
| | | Popularity Biasedness | Gangwar and Jain (2021) |

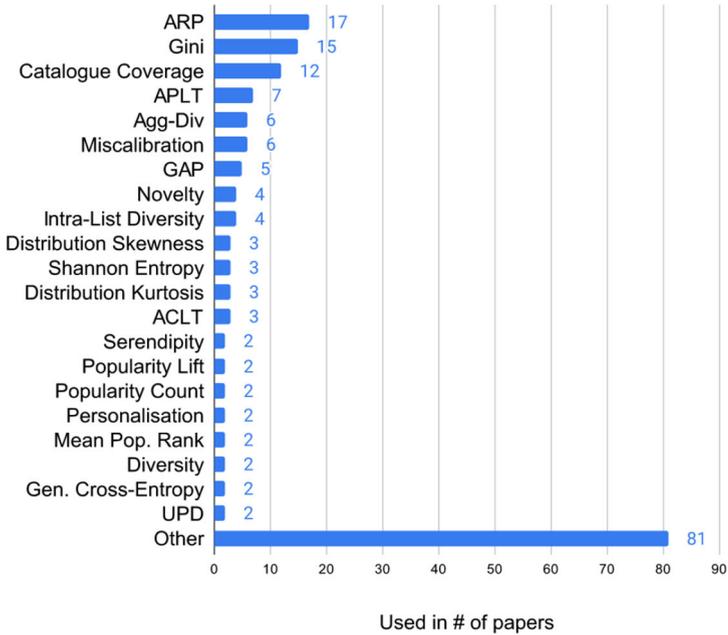

(a) Popularity and Beyond-Accuracy Metrics

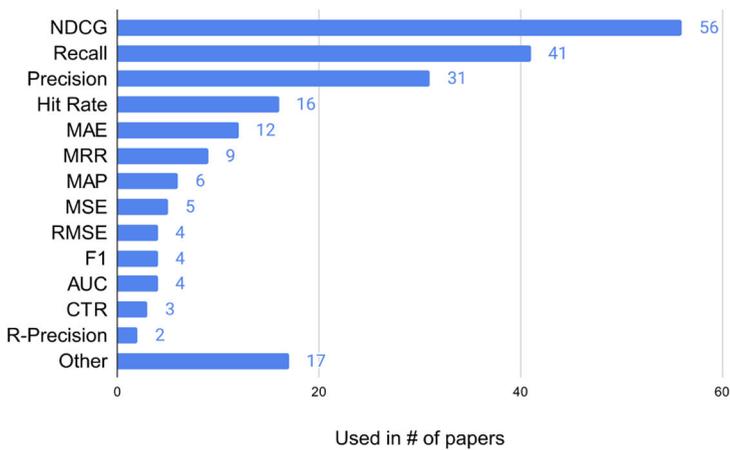

(b) Accuracy Metrics

**Fig. 12** Frequency of popularity (and other beyond-accuracy) metrics and accuracy metrics observed in the studied literature. Metrics that are used in two or more papers are explicitly included in the plot. Metrics that are used only once are grouped in the category named 'Other'

Considering the frequency of accuracy metrics, NDCG turns out to be the most commonly used one. The NDCG not only counts correctly predicted items, but also accounts for item position in a recommendation, which various other accuracy metrics do not. Other common metrics include Recall and Precision, even though they do not consider the position of the relevant items. Generally, most of the reviewed works consider one or two different accuracy metrics. In certain cases, the authors consider solely accuracy metrics for the evaluation, but measuring it separately for groups of users and items to demonstrate that recommendation quality for niche users or items can be affected by popularity bias. This is often the case for the works that mostly consider popularity bias from a purely technical standpoint and attempt to mitigate popularity bias by improving ranking/predicted ratings for tail items or serving niche-oriented users with more accurate recommendations. Generally, again a rich variety of metrics is used in the examined works. Typically, the authors would include at least one accuracy metric into the work, accompanied with more problem-specific beyond-accuracy metrics. This way it can be demonstrated that the proposed mitigation methods do not drastically affect recommendation accuracy while mitigating certain aspects of popularity bias. However, the particular choice of individual metrics is rarely discussed.

We will further discuss existing issues with common evaluation approaches and metrics in the next section.

# 7 Discussion, research gaps, and future directions

In this section, we summarize and critically discuss the findings of our analyses, and we provide an outlook of promising directions for future research.

## 7.1 Definition, applications, and datasets

Despite the significant uptake of research on the topic in the past ten years, no agreed-upon definition of what represents popularity bias has emerged so far, see our discussions of the various definitions in the literature in Sect. 2.4 and the statistics in Sect. 4.2 regarding the underlying researcher motivations to address issues of popularity bias.

Furthermore, we identified a number of research works where there was no detailed motivation provided in the papers on why popularity bias should be mitigated at all, i.e., which kinds of harm one seeks to avoid. In addition, often no explanation is provided on how the authors derived when a bias mitigation procedure is successful. In fact, even a reduction of the bias for a given metric by, e.g., 10%, might still lead to recommendations that contain many popular items.

In that context, a common assumption seems to be that recommending popular items is bad *per se*, and almost by definition leads to other effects such as limited diversity or a lack of fairness. As discussed earlier, at least for some users, the recommendation of popular items is what they expect and prefer, and some items might just be unpopular because they are of limited quality.

All in all, these observations point to a certain over-simplification of the problem and an overly abstract research operationalization, a phenomenon which can also be observed in today's research on fairness in recommender systems (Deldjoo et al. 2023). The fact that a large majority of the published research is based on datasets from the media domain, in particular on MovieLens datasets, may be seen as another factor that supports this hypothesis. In such a setting, the problem of mitigating popularity bias is reduced to designing or adopting algorithms that increase the value of certain computational bias metrics while not compromising recommendation accuracy too much. As such, popularity bias mitigation is seen to be not much different from approaches that seek to improve beyond-accuracy metrics such as diversity, novelty, or serendipity.

In practical applications, however, a more nuanced approach is required. Focusing the recommendations deliberately on popular items to some extent may in fact be a viable and successful strategy, see for example the discussions in the case of Netflix in Gomez-Uribe and Hunt (2015). In practice, two important questions in this context have to be answered: (a) when we should consider an item to be unpopular, and (b) what is the right amount of popularity bias, i.e., how do we find the right balance between recommending users what they probably like and helping them to explore new things. In many academic works on popularity bias, this balance is assumed to be given, e.g., by simply defining that the 30% least popular items are those that should be recommended more often to solve the problem.

In our work, we therefore propose a novel value- and impact-oriented definition of popularity bias, see Sect. 2.4. The main point of our definition is that popularity bias has to be addressed in case it limits the value of the recommendations or has a potentially harmful impact on some of the involved stakeholders. Adopting such a definition requires us to first think about the idiosyncrasies of the given application setting, which then allows us to select or design an appropriate computational metric. This stands in contrast to many of today's works in which the choice of the evaluation metric and of specific thresholds almost appears arbitrary. Indeed, our in-depth analysis of metrics in Sect. 6.3 showed that researchers today rely on a rich variety of application-independent, generic evaluation metrics. This leads to difficulties when comparing previous works and when trying to identify what could be considered the "state-of-the-art". As a result, this situation makes it challenging to ensure reproducibility and true progress in the area of popularity bias mitigation.

In future works, we therefore believe that application-specific considerations have to be discussed more often, ultimately leading to research work that has the potential to be more impactful in practice. One important prerequisite to enable such works however lies in the availability of additional public datasets, in particular in domains where popularity bias and the related phenomena of fairness or diversity play a central role in society.

## 7.2 Methodological issues

The indications towards an oversimplification of the problem in today's research are corroborated by our observations reported in Sect. 6 on common evaluation approaches. Almost all of today's research is based on offline experiments, which divert from the question of how users would actually perceive the value of the recommendations they receive. In this context, research on popularity bias systems suffers from a general tendency in recommender systems to rely on offline experiments (Jannach and Zanker 2021). In future works, therefore, research should be based much more often on experimental designs that include the human in the loop and which consider the impact of biased recommendations on the different stakeholders in a given application setting.

Clearly, offline experimentation will remain to have its place in research, e.g., to investigate if one algorithm has a stronger tendency to recommend popular items than another one or if popularity bias may lead to reinforcement effects in a longitudinal perspective, see, e.g., Jannach et al. (2015). Deciding whether a certain level of popularity bias is acceptable or even desirable to a certain extent however will remain to require an understanding of the specifics of a given application context. In the current literature, unfortunately no clear standards for offline evaluations have emerged yet. As discussed earlier, a variety of evaluation metrics are used and also the evaluation protocols (e.g., in terms of data splitting) can diverge significantly, again making it difficult to assess how much progress is made in the field. This problem is aggravated by the fact that the level of reproducibility in recommender systems research, and in AI in general, is still limited to a certain extent (Boratto et al. 2022; Ferrari Dacrema et al. 2021).

Putting aside specific questions of offline experiments, we argue that more impactful research on popularity bias may only be reliably achieved if we rely more often on a richer methodological repertoire in the future. This may include both alternative forms of computational experiments, e.g., simulations to study longitudinal effects, experimental designs that involve humans in the evaluation process, as well as field studies in which the effects of popularity bias are analyzed in real-world environments. Ultimately, such an approach will require us to more frequently go beyond the comparably narrow perspective of treating recommender systems research as mostly research on algorithms. Instead, it is important to adopt a more holistic research perspective, which also considers the embedding of the recommender system in a given application and the expected impact and value for the involved stakeholders. Studying phenomena such as popularity bias without considering these surrounding factors may ultimately lead to a certain stagnation in this area, leaving the question open about how impactful such research might be in practice.

## 8 Summary

Recommender systems that have a bias towards recommending mostly popular items may be of limited value both for users and for providers, and such systems may even exert harmful effects in certain application settings. In this work, we have reviewed the existing literature on popularity bias in recommender systems. This research area is currently flourishing, partly due to its relation to such important topics as fairness. Nevertheless, we found that there still exists a multitude of future directions in this area, in particular in terms of a better understanding of the real-world implications of popularity bias.

**Acknowledgements** This research was supported by industry partners and the Research Council of Norway with funding to MediaFutures: Research Centre for Responsible Media Technology and Innovation, through the Centres for Research-based Innovation scheme, project number 309339.

**Author Contributions** A.K.: Methodology; Data curation; Investigation; Writing - original draft; Visualization D.J.: Conceptualization; Methodology; Writing - original draft; Visualization M.E.: Writing - review & editing; C.T.: Writing - review & editing; Funding acquisition

**Funding** Open access funding provided by University of Bergen (incl Haukeland University Hospital)

## Declarations

**Conflict of interest** The authors declare no Conflict of interest.

**Anastasiia Klimashevskaia** is a PhD candidate at the MediaFutures Research Centre at the University of Bergen, Norway. She previously received a master's degree in Computer Science at Graz University of Technology in Graz, Austria, and a bachelor's degree in Intelligent Systems and Computational Linguistics from the Russian State University for the Humanities in Moscow, Russia. Her areas of interest have been varying greatly from robotics to media studies and natural language processing, however, Anastasiia's latest research within her PhD project has been concentrated on recommender systems and mainly the beyond-accuracy aspects of it, such as biases or fairness. She has previously published works on this topic in such international conferences as ACM RecSys and UMAP, among other experts in this field.

**Dietmar Jannach** is a Professor of Computer Science at the University of Klagenfurt, Austria. He has worked on different areas of artificial intelligence, including recommender systems, model-based diagnosis, and knowledge-based systems. He is the leading author of a textbook on recommender systems and has authored more than a hundred research papers, focusing on the application of artificial intelligence technology to practical problems.

**Mehdi Elahi** is an Associate Professor at the University of Bergen (UiB), Norway, with a Ph.D. in Computer Science from the Free University of Bozen-Bolzano (UniBz), Italy. He has published over 80 peer-reviewed papers, amassing over 3600 citations and an H-index of 29. Specializing in recommender systems, particularly for industrial applications like movie and news recommendations, he co-invented and co-owns a US patent in this area. Elahi is actively involved in the MediaFutures Research Centre, leading a work package within a 30 million euro project. He has received research credits from Amazon and Google, published influential articles in top journals and the recommender systems handbook, and organized international data challenges with companies like Spotify and XING.

**Christoph Trattner** is a full professor at the University of Bergen and the Founder and Center Director of the Research Centre for Responsible Media Technology & Innovation. He is also the founder and leader of the DARS research group at UiB. He received a PhD (with distinction), an MSc (with distinction), and a BSc in Computer Science and Telematics from Graz University of Technology (Austria). Since 2012, his research has focused on improving diets in the context of food and media. He published over 120 scientific articles in top venues in this context of research. He is a former board member of Media City Bergen, Norway's largest Media Cluster, and a member of editorial boards including Springer's Journal of Intelligent Information Systems, and AI and Ethics. Since  June 2021, he has been an appointed ACM Distinguished Speaker and ACM Senior Member.